\documentclass[10pt]{article}

\usepackage[utf8]{inputenc}
\usepackage[T1]{fontenc}
\usepackage{mathptmx}
\usepackage{microtype}
\usepackage{amsmath, amssymb}
\usepackage{graphicx}
\usepackage[dvipsnames]{xcolor}
\usepackage{booktabs}
\usepackage{longtable}
\usepackage{siunitx}
\usepackage[colorlinks=true,
            linkcolor=blue!70!black,
            citecolor=blue!70!black,
            urlcolor=blue!70!black]{hyperref}
\usepackage[nameinlink]{cleveref}
\usepackage[numbers,round,sort&compress]{natbib}
\usepackage{caption}
\usepackage{subcaption}
\usepackage[top=0.75in, bottom=1in, left=0.75in, right=0.75in, letterpaper]{geometry}
\graphicspath{{figures/}}

\begin{document}

\twocolumn[{%
  \centering
  {\LARGE\bfseries HistoAtlas: A Pan-Cancer Morphology Atlas Linking\\ Histomics to Molecular Programs and Clinical Outcomes\par}
  \vspace{1em}
  {\normalsize Pierre-Antoine Bannier\par}
  \vspace{0.8em}
  \rule{\textwidth}{0.4pt}
  \vspace{0.8em}
}]

{\small\bfseries

We present HistoAtlas, a pan-cancer computational atlas that extracts 38
interpretable histomic features from 6,745 diagnostic H\&E slides across
21 TCGA cancer types and systematically links every feature to survival,
gene expression, somatic mutations, and immune subtypes. All associations
are covariate-adjusted, multiple-testing corrected, and classified into
evidence-strength tiers. The atlas recovers known biology, from immune
infiltration and prognosis to proliferation and kinase signaling, while
uncovering compartment-specific immune signals and morphological subtypes
with divergent outcomes. Every result is
spatially traceable to tissue compartments and individual cells, statistically
calibrated, and openly queryable. HistoAtlas enables systematic, large-scale biomarker discovery from
routine H\&E without specialized staining or sequencing.
Data and an interactive web atlas are freely available at
\url{https://histoatlas.com}.
}

\vspace{0.5em}

\noindent{\small\textbf{Correspondence:} \href{mailto:pierreantoine.bannier@gmail.com}{pierreantoine.bannier@gmail.com}}
\vspace{0.5em}


\section{Introduction}
\label{sec:intro}

Histopathological examination of hema\-toxylin-and-eosin-stained (H\&E) tissue sections
remains the gold standard for cancer diagnosis~\cite{bera2019, kather2020pancancer}. Every diagnostic slide encodes
quantitative information, from cell densities and nuclear morphology to spatial
organization of immune infiltrates and stromal architecture~\cite{gurcan2009}.
In existing pan-cancer resources, this information is collapsed into
categorical grades or discarded entirely~\cite{vanderlaak2021}.
Genomics~\cite{kandoth2013}, transcriptomics~\cite{hoadley2014}, proteomics~\cite{li2023cptac}, and epigenomics~\cite{corces2018} each have mature
pan-cancer resources that enable systematic cross-cancer comparison.
Yet, histopathology, the most routinely generated cancer data modality, lacks
an equivalent quantitative atlas.

The Cancer Genome Atlas (TCGA) established the paradigm for
multi-omic integration across cancer types, cataloging somatic mutations,
copy-number alterations, gene expression programs, and epigenetic
landscapes~\cite{hoadley2018}. Thorsson et~al.\ extended this framework
to immune biology, defining six immune subtypes that stratify prognosis across 33
cancer types using transcriptomic and genomic
features~\cite{thorsson2018immune}. Nonetheless, neither resource incorporates
quantitative morphological data. This is a notable omission because the spatial context of immune infiltration carries prognostic information independent of molecular subtyping, as formalized in the Immunoscore~\cite{galon2006, pages2018, fridman2012immune}.
Saltz et~al.\ mapped bulk tumor infiltrating lymphocyte (TIL) density across 13 TCGA cancer types from deep-learning spatial maps~\cite{saltz2018til}, demonstrating the feasibility of pan-cancer morphological analysis from H\&E. However, their approach reports a single bulk density score without compartment-specific resolution or linkage to gene expression programs.

Computational pathology has made rapid progress in extracting quantitative features from digitized slides~\cite{bera2019}.
Early morphometric studies demonstrated that automated image features carry prognostic value in individual cancer types~\cite{beck2011, yu2016, cooper2012}.
Deep-learning classifiers now predict molecular alterations~\cite{echle2021, coudray2018, bannier2024}, microsatellite instability~\cite{kather2019msi, saillard2023msi}, gene expression~\cite{schmauch2020he2rna}, and survival~\cite{kather2019treatment} directly from H\&E with high accuracy.
More recently, foundation models such as UNI~\cite{chen2024uni}, Virchow~\cite{vorontsov2024virchow}, or H0~\cite{hoptimus0}, trained on large datasets of pan-tumor tissue via self-supervised learning, produce information-rich slide embeddings.
Yet, these embeddings do not readily decompose into interpretable biological features such as cell densities, spatial distances, or tissue compartment fractions~\cite{vanderlaak2021}.
In response to this interpretability gap, several groups have proposed explicit feature-based representations: Diao et al.~\cite{diao2021hif} combined cell- and tissue-level predictions into hundreds of human-interpretable descriptors, and Abel et al.~\cite{abel2024} derived large collections of nuclear morphometric features linked to genomic instability and prognosis. These studies show that interpretable H\&E features carry rich biological signal, but their emphasis on large feature spaces does not naturally organize into a concise morphology atlas grounded in a small set of reproducible, compartment-resolved features.

Public resources mirror this gap: cBioPortal provides molecular data without morphology~\cite{cerami2012cbio}, TCIA hosts raw slides without precomputed features~\cite{clark2013tcia}, and the Human Protein Atlas maps protein expression without quantitative morphometrics~\cite{uhlen2015hpa}.
These gaps leave cancer researchers without a resource that bridges
morphology and molecular biology at pan-cancer scale. Such a resource would need to
combine interpretable histomic features with systematic molecular linkage across cancer types, explicit multiple-testing control, and traceability from statistical associations back to tissue compartments and individual cells.

Here we present HistoAtlas, a pan-cancer morphology atlas built from \num{38}
quantitative histomic features extracted from \num{6745} TCGA diagnostic slides across 21
cancer types (plus a pooled pan-cancer analysis). We systematically test every feature for association with survival,
gene expression, mutations, copy-number variation, and immune subtypes with
explicit correction families and evidence-strength badges (strong, moderate, suggestive, or insufficient).
All results are released as a web atlas in which every association is spatially traceable to specific tissue compartments and individual cells (Fig.\,\ref{fig:fig6}).
We demonstrate that resolving immune cells by tissue compartment uncovers a stronger protective observational association between intratumoral lymphocyte density and survival than its stromal counterpart, a distinction diluted in bulk H\&E-derived TIL scoring approaches. Among morphologically distinct clusters, morphology separates quiescent from hormone-driven subgroups with divergent outcomes.


\section{Results}
\label{sec:results}

\subsection{A quantitative atlas of cancer morphology}
\label{sec:results:atlas}

We constructed HistoAtlas from \num{6745} H\&E-stained diagnostic slides
spanning 21 TCGA solid-tumor cancer types (Supplementary Table~3).
Twelve additional cancer types were excluded because their dominant cell
morphologies (lymphoid, glial, melanocytic, mesenchymal, neuroendocrine,
renal tubular, or germ cell) fall outside the training domain of the
segmentation models (Supplementary Table~3).

Two automated segmentation stages converted whole-slide images into
quantitative measurements (\S\ref{sec:methods:features}).
First, a UNet-based tissue segmentation model classified approximately
1.4~m$^2$ of tissue into five compartments (tumor [mean 44.9\% of tissue
area], stroma [45.4\%], necrosis, blood, and normal epithelium; Fig.~\ref{fig:fig1}a), with
tumor and stroma together accounting for over 90\% of the analyzed area.
Second, the HistoPLUS cell detection and classification
model~\cite{adjadj2025histoplus} identified more than 4.4~billion individual
cells belonging to nine types: tumor cells, lymphocytes, fibroblasts,
neutrophils, eosinophils, plasmocytes, apoptotic bodies, mitotic figures,
and red blood cells.

From these segmentations we derived 38 histomic features organized into
five categories: tissue composition, cell densities, nuclear morphology and
kinetics, spatial organization, and spatial heterogeneity
(definitions in Supplementary Table~1;
descriptive statistics in Supplementary Table~\ref{tab:feature_stats};
preprocessing in \S\ref{sec:methods:preprocessing}).

We then tested each feature for associations with survival and molecular
programs across all 22~cohorts.
For survival, we fitted Cox proportional-hazards models for each
combination of 38~features, 22~cohorts (21~cancer types plus a pan-cancer
cohort), and four endpoints (overall, disease-specific, disease-free, and
progression-free survival), yielding \num{5623} evaluable associations
(of a theoretical maximum of \num{6688}; the remainder were excluded for
insufficient sample size or events) under two
adjustment tiers, unadjusted and adjusted for age, sex, stage, and
tissue source site (\S\ref{sec:methods:survival}).
After Benjamini--Hochberg correction within predefined correction families
(\S\ref{sec:methods:multiple_testing}; Supplementary Table~6), 260
associations were significant at a false discovery rate of 0.05.
All 260 passed the proportional-hazards assumption (Schoenfeld
$P \geq 0.05$; Supplementary Table~7), because associations with
PH violations have their Cox $P$-values invalidated before
BH correction (\S\ref{sec:methods:survival}); restricted mean survival time (RMST) summaries
are provided as complementary measures for all associations.

For molecular associations, we computed \num{487638} Spearman rank
correlations between 38~histomic features and 293~molecular targets,
comprising 133 curated cancer genes assessed for both mRNA expression and
copy-number variation (Supplementary Table~4), 21~Hallmark pathway
activity scores (of the 50 Hallmark gene sets, 21 had sufficient matched data), and 6~immune cell-fraction scores, across 22~cohorts
under two adjustment tiers (\S\ref{sec:methods:correlations}).
After family-wise Benjamini--Hochberg correction
(Supplementary Table~6), \num{88920} correlations (18.2\%) were
significant at a false discovery rate of 0.05, with the highest yield among immune cell
fractions (39.2\%), pathway scores (30.4\%), and gene expression (24.9\%),
and the lowest among copy-number variation (6.3\%)
(Supplementary Table~\ref{tab:correlation_breakdown}).
Sample sizes vary across analyses because not all slides have matched
molecular or clinical annotation; exact counts are reported per analysis
throughout.
The following subsections present what these associations show,
beginning with a pan-cancer morphological landscape and progressing
to compartment-resolved survival signals.

\subsection{The pan-cancer morphological landscape recovers canonical biology}
\label{sec:results:landscape}

Our pipeline extracts 38 quantitative histomic features from each diagnostic
H\&E slide through automated tissue segmentation, cell detection, and
spatial analysis (Fig.~\ref{fig:fig1}a).
Pairwise Spearman correlation across all \num{6745} slides revealed structured
feature modules -- density features form a tight positive-correlation block,
morphology features cluster together, and cross-module anti-correlations
delineate distinct biological axes (Fig.~\ref{fig:fig1}b) -- confirming that
the 38 features capture complementary aspects of tissue biology.
To visualize the morphological landscape, we projected all \num{6745} slides
into a two-dimensional UMAP embedding computed from these features
(\S\ref{sec:methods:clustering}; Fig.~\ref{fig:fig1}c).
Cancer types occupied distinct regions of the embedding, with morphologically
related types positioned adjacently: squamous carcinomas (HNSC, LUSC, CESC)
clustered in a region of elevated nuclear pleomorphism, while hormone-driven
adenocarcinomas (BRCA, PRAD) occupied a low-proliferation region.
Unsupervised K-means clustering of the z-scored feature vector, without any
molecular input, yielded 10 pan-cancer (L1) clusters ($K$ selected by inspection of silhouette,
Calinski--Harabasz, Davies--Bouldin, and gap statistic metrics;
\S\ref{sec:methods:clustering}; Fig.~\ref{fig:fig1}d,e) and 69 cancer-specific (L2)
subclusters.
Bootstrap stability analysis (50 iterations, 80\% subsamples) confirmed robust
cluster assignments (mean adjusted Rand index~$= 0.72$, Jaccard~$= 0.81$).
The adjusted Rand index between L1 clusters and cancer-type labels was 0.15,
confirming that the clusters capture morphological variation that is not
reducible to cancer-type identity.

Pathway and immune subtype enrichment analysis revealed that these purely
morphological clusters align with canonical molecular programs
(\S\ref{sec:results:archetypes}).
All pathway enrichments below are Cliff's $\delta$ computed on
Hallmark gene set scores~\cite{liberzon2015} (Supplementary Table~\ref{tab:pathway_signatures}).
Cluster~4 (76\% THYM) exhibited strong immune rejection pathway enrichment
($\delta = 0.67$, 95\% CI $[0.60, 0.73]$,
$P_{\mathrm{adj}} = 3.3 \times 10^{-40}$), consistent with the
active T-cell maturation environment that defines thymic
biology~\cite{radtke2013, radovich2018thymoma}.
Cluster~6 (61\% COAD and READ) showed dominant Wnt/$\beta$-catenin
signaling ($\delta = 0.46$, 95\% CI $[0.42, 0.50]$,
$P_{\mathrm{adj}} = 1.3 \times 10^{-82}$) and C1 wound-healing immune
subtype enrichment (OR~$= 5.59$, 95\% CI $[4.69, 6.67]$,
$P_{\mathrm{adj}} = 1.1 \times 10^{-88}$), recapitulating the constitutive
WNT activation that characterizes colorectal
tumorigenesis~\cite{tcga2012crc}.
Cluster~8 (44\% BRCA, 24\% PRAD) displayed estrogen response upregulation
($\delta = 0.52$, 95\% CI $[0.49, 0.56]$,
$P_{\mathrm{adj}} = 2.2 \times 10^{-160}$) and proliferation suppression
($\delta = -0.51$, 95\% CI $[-0.54, -0.48]$,
$P_{\mathrm{adj}} = 3.8 \times 10^{-154}$), consistent with the
hormone-driven, genomically quiet phenotype of luminal breast and prostate
cancers~\cite{perou2000, hoadley2018}.
The algorithm received no molecular input, yet grouped thymomas by immune
rejection pathways, colorectal cancers by WNT activation, and hormone-driven tumors
by estrogen response.

Because L1 clusters dominated by a single cancer type could trivially inherit
that type's molecular profile, we examined two additional lines of evidence.
First, Cluster~3 ($n = \num{1012}$) spans five cancer types with no dominant
contributor (HNSC 17.7\%, STAD 17.2\%, BLCA 14.3\%, LUSC 14.3\%, LUAD
11.8\%) yet showed coherent enrichment for hypoxia
($\delta = 0.34$, 95\% CI $[0.30, 0.38]$,
$P_{\mathrm{adj}} = 1.7 \times 10^{-57}$), interferon-$\gamma$ response
($\delta = 0.46$, 95\% CI $[0.43, 0.49]$,
$P_{\mathrm{adj}} = 1.3 \times 10^{-105}$), and C2 (IFN-$\gamma$ dominant)
immune subtype (OR~$= 2.74$, 95\% CI $[2.37, 3.17]$,
$P_{\mathrm{adj}} = 2.1 \times 10^{-40}$).
Second, within-cancer (L2) subclusters showed biology beyond cancer-type
identity: within BRCA alone, subcluster~2 ($n = 280$) was enriched for C2
immune subtype (OR~$= 4.86$, 95\% CI $[3.58, 6.59]$,
$P_{\mathrm{adj}} = 8.1 \times 10^{-25}$) and interferon-$\gamma$ response
($\delta = 0.47$, 95\% CI $[0.39, 0.54]$,
$P_{\mathrm{adj}} = 3.8 \times 10^{-28}$), while subcluster~3 ($n = 325$)
showed estrogen response enrichment ($\delta = 0.26$, 95\% CI $[0.19, 0.34]$,
$P_{\mathrm{adj}} = 1.2 \times 10^{-9}$) and depletion across all six immune
pathways.
These within-cancer results confirm that the histomic features capture
biological heterogeneity not reducible to cancer-type identity.
The remaining clusters and their survival associations are detailed in
\S\ref{sec:results:archetypes}.

\subsection{Spatial immune topology is associated with survival in a compartment-specific manner}
\label{sec:results:immune}

Unlike bulk TIL scoring approaches~\cite{salgado2015til, saltz2018til},
HistoAtlas quantifies immune cell density, spatial proximity, and
infiltration patterns separately in the intratumoral, stromal, and invasive
front compartments. All survival associations in this subsection use
Cox regression adjusted for age, sex, stage, and stratified by tissue
source site for overall survival
(\S\ref{sec:methods:survival}; pan-cancer models additionally stratified
by cancer type).

Pan-cancer analysis revealed compartment-specific differences in prognostic
strength (Fig.~\ref{fig:fig2}a). Intratumoral lymphocyte density was
associated with favorable outcomes (pan-cancer hazard ratio
[HR]~$= 0.87$, 95\% CI $[0.81, 0.93]$,
$P_{\mathrm{adj}} = 9.8 \times 10^{-4}$, $n = 4{,}560$), whereas stromal
lymphocyte density showed a weaker, attenuated protective effect
(HR~$= 0.89$, 95\% CI $[0.83, 0.97]$,
$P_{\mathrm{adj}} = 0.031$, $n = 4{,}561$). Intratumoral
lymphocyte density showed a protective direction (HR~$< 1$) in 11 of 17
evaluable cancer types, with BRCA exhibiting the strongest effect
(HR~$= 0.72$, 95\% CI $[0.60, 0.88]$,
$P_{\mathrm{adj}} = 0.018$, $n = 960$; Fig.~\ref{fig:fig2}b)
followed by HNSC (HR~$= 0.74$, 95\% CI $[0.63, 0.87]$,
$P_{\mathrm{adj}} = 3.9 \times 10^{-3}$, $n = 444$). In BRCA, stromal
lymphocyte density showed a weaker, non-significant association
(HR~$= 0.93$, 95\% CI $[0.77, 1.12]$, $P_{\mathrm{adj}} = 0.67$),
indicating that the intratumoral compartment carries the dominant
prognostic signal (Fig.~\ref{fig:fig6}b). Aggregate TIL scores that combine both compartments
dilute this compartment-specific effect.

Spatial proximity features provided an additional prognostic axis.
Tumor-lymphocyte nearest-neighbor distance at the invasive front, a spatial
measure of immune exclusion~\cite{joyce2015exclusion, chen2017oncology}, inversely correlated
with CD8A expression in BRCA ($\rho = -0.53$,
$P_{\mathrm{adj}} = 1.8 \times 10^{-68}$, $n = 958$; Fig.~\ref{fig:fig2}c).

Gene-level correlations validated the biological identity of these features.
In BRCA, intratumoral lymphocyte density correlated with cytotoxic T-cell
markers and immune checkpoint genes (CD8A: $\rho = 0.59$, 95\% CI
$[0.54, 0.63]$; TIGIT: $\rho = 0.63$, 95\% CI $[0.59, 0.67]$; both
$P_{\mathrm{adj}} < 10^{-85}$, $n = 958$; Fig.~\ref{fig:fig2}d). These features also
discriminated Thorsson immune subtypes~\cite{thorsson2018immune}:
peritumoral immune richness (the number of distinct immune cell types
detected within 50~\textmu m of the tumor boundary; Supplementary Table~1)
explained 13\% of immune subtype variance (Kruskal--Wallis
$\eta^2 = 0.13$, 95\% CI $[0.12, 0.15]$,
$P_{\mathrm{adj}} = 3.0 \times 10^{-159}$, $n = 5{,}590$; pan-cancer),
consistent with concordance between histomic and transcriptomic immune
classifications.
A composite feature, interface-normalized immune pressure (lymphocyte count
within 50~\textmu m of the tumor--stroma boundary divided by interface
length, cells\,mm$^{-1}$; Supplementary Table~1), was protective in HNSC
(HR~$= 0.74$, 95\% CI $[0.63, 0.86]$,
$P_{\mathrm{adj}} = 3.9 \times 10^{-3}$, $n = 444$; a value similar to intratumoral lymphocyte density, reflecting the high correlation between these features).

Additional features showed consistent protective trends across cancer types.
Lymphocyte density spatial heterogeneity was protective in 14 of 17 evaluable
cancer types (unadjusted model). The unadjusted
associations for interface-normalized immune pressure in BRCA
(HR~$= 0.72$, $P_{\mathrm{adj}} = 7.6 \times 10^{-3}$) and LIHC
(HR~$= 0.79$, $P_{\mathrm{adj}} = 0.038$) did not survive covariate
adjustment.

\subsection{Morphometric features encode molecular programs}
\label{sec:results:molecular}

We next tested whether purely morphometric features serve as proxies
for molecular programs. Of the \num{487638} histomic--molecular
correlations (\S\ref{sec:results:atlas}), \num{88920} (18.2\%) were
significant at FDR~$< 0.05$.
Under a permutation null model (100 shuffles of molecular labels within each cancer type, with per-cancer-type BH correction matching the production pipeline), 0\% of pairs were significant at the same threshold, confirming that the observed 18.2\% discovery rate reflects genuine biological signal rather than statistical artifact (Supplementary Methods).
The correlation
structure was biologically coherent: immune density features correlated
with immune pathway signatures, proliferation features with cell cycle
pathways, and invasion features with epithelial-mesenchymal transition
(EMT) scores (Fig.~\ref{fig:fig3}a).
Among significant pairs, the median
absolute $\rho$ was 0.18 (IQR 0.13--0.27).
Fig.~\ref{fig:fig3}b shows the distribution of effect sizes for
pan-cancer adjusted-model associations, stratified by molecular data
type: gene expression (\num{4371}/\num{5453} significant, 80\%),
Hallmark pathways (\num{1692}/\num{2050}, 83\%), and copy-number
variation (\num{2845}/\num{5453}, 52\%).

Three examples from breast cancer (BRCA, $n = 958$; unadjusted model)
illustrate the strength of this
morphology-to-molecular correspondence.
First, mitotic index correlated with canonical proliferation markers
(PLK1: $\rho = 0.56$, 95\% CI $[0.51, 0.61]$,
$P_{\mathrm{adj}} = 5.2 \times 10^{-77}$;
additional markers including AURKA, MKI67, CCNB1, and TOP2A).
Second, invasion depth showed modest correlations
($|\rho| = 0.25$--$0.32$) consistent with the classical EMT
axis~\cite{nieto2016emt}, with ZEB1 as the strongest correlate
($\rho = 0.32$, 95\% CI $[0.26, 0.37]$,
$P_{\mathrm{adj}} = 8.4 \times 10^{-23}$) and an inverse correlation
with the epithelial marker CDH1 ($\rho = -0.25$, 95\% CI $[-0.32,
-0.19]$, $P_{\mathrm{adj}} = 1.0 \times 10^{-14}$).
Third, nuclear pleomorphism anti-correlated with luminal
differentiation markers (BCL2: $\rho = -0.37$, 95\% CI $[-0.43,
-0.32]$, $P_{\mathrm{adj}} = 8.1 \times 10^{-32}$; ESR1:
$\rho = -0.36$, 95\% CI $[-0.41, -0.30]$,
$P_{\mathrm{adj}} = 4.4 \times 10^{-29}$), consistent with the
histological grading criteria of Elston and
Ellis~\cite{elston1991grading}.
The mitotic index--PLK1 correspondence generalized across cancer types
(LUAD: $\rho = 0.65$, $n = 437$; LIHC: $\rho = 0.60$, $n = 348$;
pan-cancer: $\rho = 0.68$, $n = 5{,}875$;
all $P_{\mathrm{adj}} < 10^{-27}$).

Invasion depth also inversely correlated with cell cycle pathway
scores in BRCA ($\rho = -0.30$, 95\% CI $[-0.36, -0.25]$,
$P_{\mathrm{adj}} = 4.6 \times 10^{-21}$, $n = 957$).
This slide-level inverse association between invasion and
proliferation is consistent with the ``go-or-grow''
hypothesis~\cite{giese2003glioma, hatzikirou2012}, although it cannot establish
single-cell-level mutual exclusivity.
Together, these correspondences confirm that histomic features
capture interpretable aspects of known biological programs, providing
a morphology-to-molecular bridge that operates without specialized
staining or sequencing.

\subsection{Morphological clusters define molecular archetypes}
\label{sec:results:archetypes}

Beyond the pathway enrichments that independently recovered canonical
biology (\S\ref{sec:results:landscape}), the 10 L1 clusters also carried
distinct mutational and immune subtype profiles (Fig.~\ref{fig:fig4}a,b).
Mutation enrichment analysis (Fisher's exact test, FDR~$< 0.05$) showed
Cluster~6 (61\% CRC) enriched for TTN (odds ratio [OR]~$= 1.91$, 95\% CI
$[1.58, 2.31]$, $P_{\mathrm{adj}} = 1.4 \times 10^{-9}$), FAT4
(OR~$= 1.91$, 95\% CI $[1.46, 2.49]$,
$P_{\mathrm{adj}} = 4.9 \times 10^{-5}$), and SYNE1 (OR~$= 1.86$,
95\% CI $[1.47, 2.35]$,
$P_{\mathrm{adj}} = 1.1 \times 10^{-5}$), mutations frequently observed
in colorectal genomes. Cluster~8 (44\% BRCA, 24\%
PRAD) was depleted for chromatin modifier mutations (KMT2D OR~$= 0.42$,
95\% CI $[0.31, 0.55]$,
$P_{\mathrm{adj}} = 5.8 \times 10^{-11}$; ZFHX4 OR~$= 0.44$, 95\% CI
$[0.33, 0.58]$, $P_{\mathrm{adj}} = 4.8 \times 10^{-10}$), consistent
with a genomically quiet, hormone-driven phenotype. Because cluster
molecular enrichments partly reflect cancer-type composition (e.g.,
Cluster~4 is 76\% THYM), within-cancer-type (L2) enrichments that control
for this confound are available in the web atlas.

\begin{figure*}[t]
  \centering
  \includegraphics[width=\textwidth]{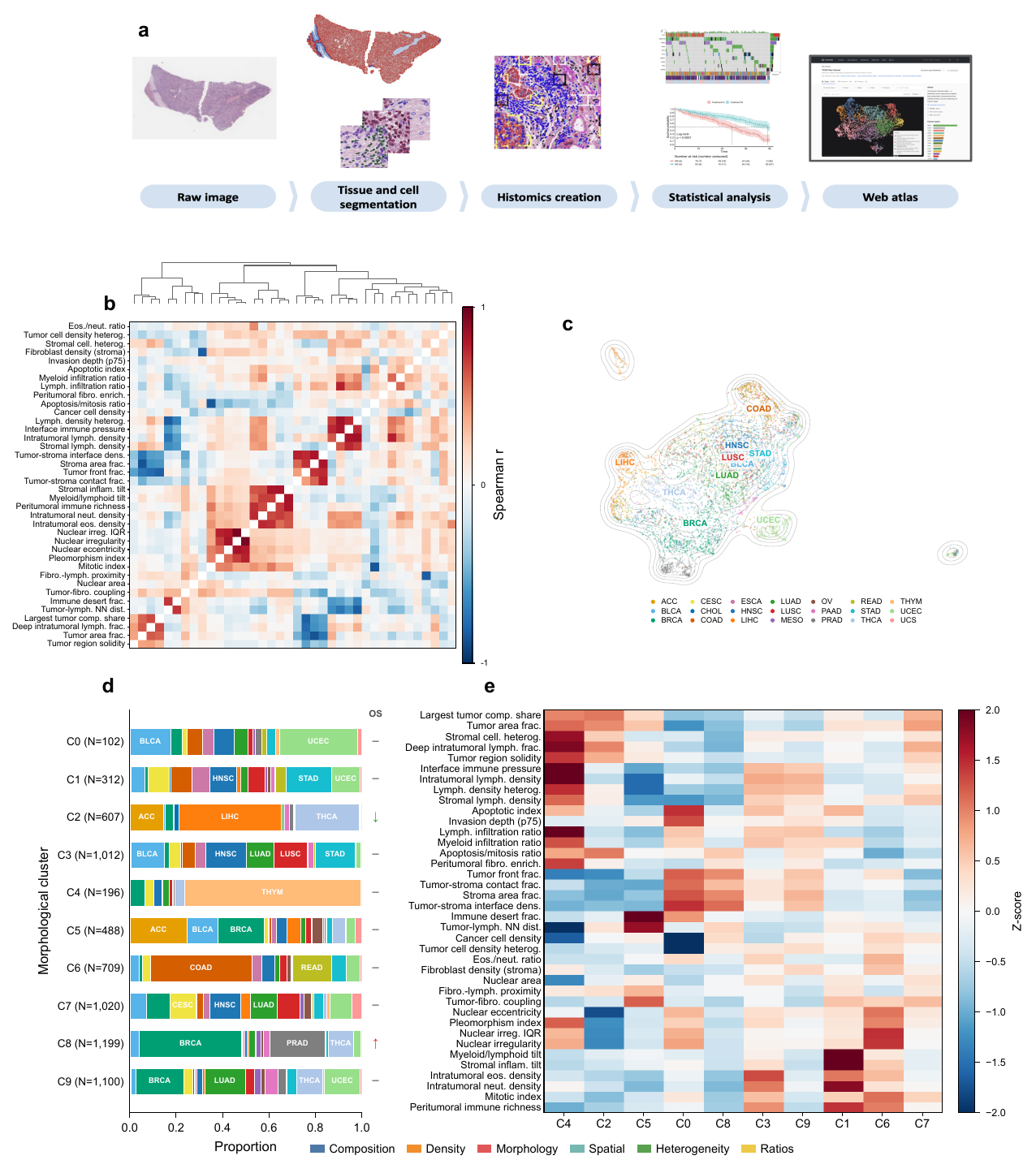}
  \caption{%
    \textbf{The HistoAtlas pipeline and pan-cancer morphological landscape.}
    (a)~Overview of the computational pipeline. Diagnostic H\&E-stained whole-slide images (\num{6745} slides, 21 TCGA cancer types) are segmented into tissue compartments (tumor core, stroma, invasive front), followed by cell-level detection and classification of 9 cell types. From each slide, 38 quantitative histomic features are extracted spanning tissue composition, cell densities, nuclear morphology, spatial immune topology, microenvironment heterogeneity, and cell-type ratios.
    (b)~Pairwise Spearman correlation matrix of the 38 features computed across all \num{6745} slides. Ward-linkage hierarchical clustering reveals structured modules: density features form a tight positive-correlation block, morphology features cluster together, and cross-module anti-correlations delineate distinct biological axes. Left color bar indicates feature category. Diagonal entries are masked.
    (c)~UMAP embedding of all \num{6745} slides colored by cancer type. Cancer types with distinct morphological programs (e.g., THYM, THCA) occupy separated regions, while adenocarcinomas (BRCA, LUAD, STAD) partially overlap. Gray contour lines indicate point density.
    (d)~Cancer type composition of each L1 morphological cluster ($K = 10$, horizontal stacked bars), with cluster sizes indicated at left. Cancer types constituting more than 10\% of a cluster are labeled within the bar. Right annotation shows overall survival direction per cluster (green arrow: significantly protective, HR~$< 1$; red arrow: significantly adverse, HR~$> 1$; gray dash: non-significant).
    (e)~Heatmap of z-scored mean feature values per cluster, with Ward-linkage hierarchical clustering applied to both features (rows) and clusters (columns). Feature labels are colored by category. Red indicates elevated values; blue indicates suppressed values relative to the pan-cancer mean. Values are clipped at $z = \pm 2$ for visualization.
    $N = \num{6745}$ slides from 21 TCGA cancer types.
  }
  \label{fig:fig1}
\end{figure*}

Cluster-level survival analysis used Cox regression stratified by
cancer type (Fig.~\ref{fig:fig1}d). This analysis revealed a
prognostically important distinction among morphologically distinct
clusters. Cluster~2
($n = 607$; 44\% LIHC, 28\% THCA) displayed profoundly quiescent
morphology: proliferation pathway scores were suppressed relative to all
other slides (Cliff's $\delta = -0.58$,
$P_{\mathrm{adj}} = 4.2 \times 10^{-102}$; E2F targets), and it showed
favorable survival (HR~$= 0.54$, 95\% CI $[0.40, 0.73]$,
$P_{\mathrm{adj}} = 6.3 \times 10^{-4}$; $n = 516$ with events).
Cluster~5 ($n = 488$; 25\% ACC, 20\% BRCA) showed immune spatial
exclusion (depleted cytotoxic immune activity; Cliff's
$\delta = -0.54$, $P_{\mathrm{adj}} = 2.3 \times 10^{-62}$;
allograft rejection) with near-average proliferative activity, and a
non-significant adverse trend (HR~$= 1.17$, 95\% CI $[0.96, 1.42]$,
$P_{\mathrm{adj}} = 0.28$).

Thorsson immune subtype~\cite{thorsson2018immune} composition further
distinguished the two clusters. Cluster~5 was enriched for C4
(lymphocyte depleted; OR~$= 5.49$, 95\% CI $[4.23, 7.11]$,
$P_{\mathrm{adj}} = 2.8 \times 10^{-30}$; 28\% of slides) and depleted
for C2 (IFN-$\gamma$ dominant; OR~$= 0.48$,
$P_{\mathrm{adj}} = 3.9 \times 10^{-8}$). Cluster~2 showed combined C4
(OR~$= 7.14$, 95\% CI $[5.72, 8.93]$,
$P_{\mathrm{adj}} = 1.3 \times 10^{-56}$) and C3 (inflammatory;
OR~$= 4.99$, 95\% CI $[4.13, 6.02]$,
$P_{\mathrm{adj}} = 1.6 \times 10^{-60}$) enrichment (85\% combined;
Fig.~\ref{fig:fig4}a). Although C3 is labeled ``inflammatory,'' Cluster~2's
morphology was uniformly quiescent, with suppressed lymphocyte density and
proliferative indices, suggesting that its C3-classified tumors represent a
quiescent inflammatory state rather than active immune engagement. Immune
subtype labels alone classified both clusters as immune-depleted variants
but did not distinguish their divergent proliferative states; the
morphological axis of quiescent-cold versus hormone-driven tumors added
prognostic information that transcriptomic subtyping did not capture.

Cluster~8 (BRCA/PRAD, hormone-driven) showed adverse
survival (HR~$= 1.37$, 95\% CI $[1.15, 1.62]$,
$P_{\mathrm{adj}} = 1.8 \times 10^{-3}$, $n = 1{,}112$).
The remaining clusters did not reach significance after BH correction.
Hazard ratios and $P$-values for all 10 clusters are shown in
Fig.~\ref{fig:fig1}d.

\subsection{Reporting what the atlas detects and what it cannot}
\label{sec:results:stats}

HistoAtlas accompanies every association with
Benjamini--Hochberg-corrected~\cite{benjamini1995} $P$-values, bootstrap
confidence intervals, effect sizes, and evidence-strength badges
(Fig.~\ref{fig:fig5}; statistical details in
\S\ref{sec:methods:multiple_testing}--\ref{sec:methods:power}).

We assessed batch effects from tissue source site (TSS) using principal
variance component analysis
(PVCA)~\cite{li2009pvca} and silhouette scores. At the pan-cancer level,
PVCA attributed 44.7\% of feature variance to TSS, 32.7\% to cancer-type
identity, and 22.6\% to residual (Fig.~\ref{fig:fig5}a). Because TSS is
partially confounded with cancer type (most sites contribute primarily one
cancer type), the 44.7\% TSS component includes both genuine institutional
variation and cancer-type-associated morphological differences. Within
individual cancer types, where batch effects could confound feature--outcome
associations, per-cancer-type batch variance ranged from 2.7\% (ACC) to
29.1\% (ESCA), and all 20 per-cancer silhouette scores by TSS were negative
(range $-0.18$ to $-0.0004$; one cancer type [CHOL] was excluded from
per-cancer batch QC due to insufficient TSS diversity), indicating that no
cancer type exhibited TSS-driven sub-clustering.

Spearman correlation $P$-values use the analytical $t$-test approximation
($t = r\sqrt{\mathrm{df}/(1-r^2)}$;
\S\ref{sec:methods:correlations}), validated by near-perfect concordance
with a permutation-based reference ($\rho > 0.999$, $n = \num{282278}$
pairs). As a calibration check, we verified that raw
$P$-values for the weakest-signal features (those with median effect size in
the bottom quartile) followed an approximately uniform distribution,
consistent with the null expectation.

To quantify what the atlas \emph{cannot} detect, we computed the minimum
detectable effect size (MDES) at 80\% power for every analysis, using the
Schoenfeld--Freedman
approximation~\cite{schoenfeld1983, freedman1982} for survival associations
and the Fisher $z$-transform for correlations. MDES varies across cancer
types because sample sizes and event counts differ: well-powered cancer types
such as BRCA ($n = 960$ for OS, 135 events) can detect hazard ratios as
small as 1.62, whereas underpowered types such as cholangiocarcinoma (CHOL,
$n = 36$) require hazard ratios exceeding 3.75 (Fig.~\ref{fig:fig5}b).

Each association receives an evidence-strength badge (strong, moderate,
suggestive, or insufficient) computed from adjusted $P$-value, effect size
magnitude, confidence interval width, and sample size
(\S\ref{sec:methods:power}). Across \num{5623} survival associations
(38~features $\times$ 22~cohorts $\times$ 4~endpoints $\times$
2~adjustment tiers, excluding combinations with insufficient data),
33 (0.6\%) achieved strong evidence, 167 (3.0\%) moderate, 577 (10.3\%)
suggestive, and \num{4846} (86.2\%) insufficient.
The predominance of insufficient evidence reflects the limited statistical
power of smaller cohorts: most insufficient-evidence pairs involve cancer
types with $n < 100$, where MDES exceeds clinically meaningful thresholds
(Fig.~\ref{fig:fig5}b). The 33 strong and 167 moderate associations span
multiple cancer types and all five feature categories, providing a curated
set of high-confidence findings.
Cross-endpoint replication rates for DSS, DFS, and PFS are discussed in Supplementary Note~2.


\begin{figure*}[p]
  \centering
  \includegraphics[width=\textwidth]{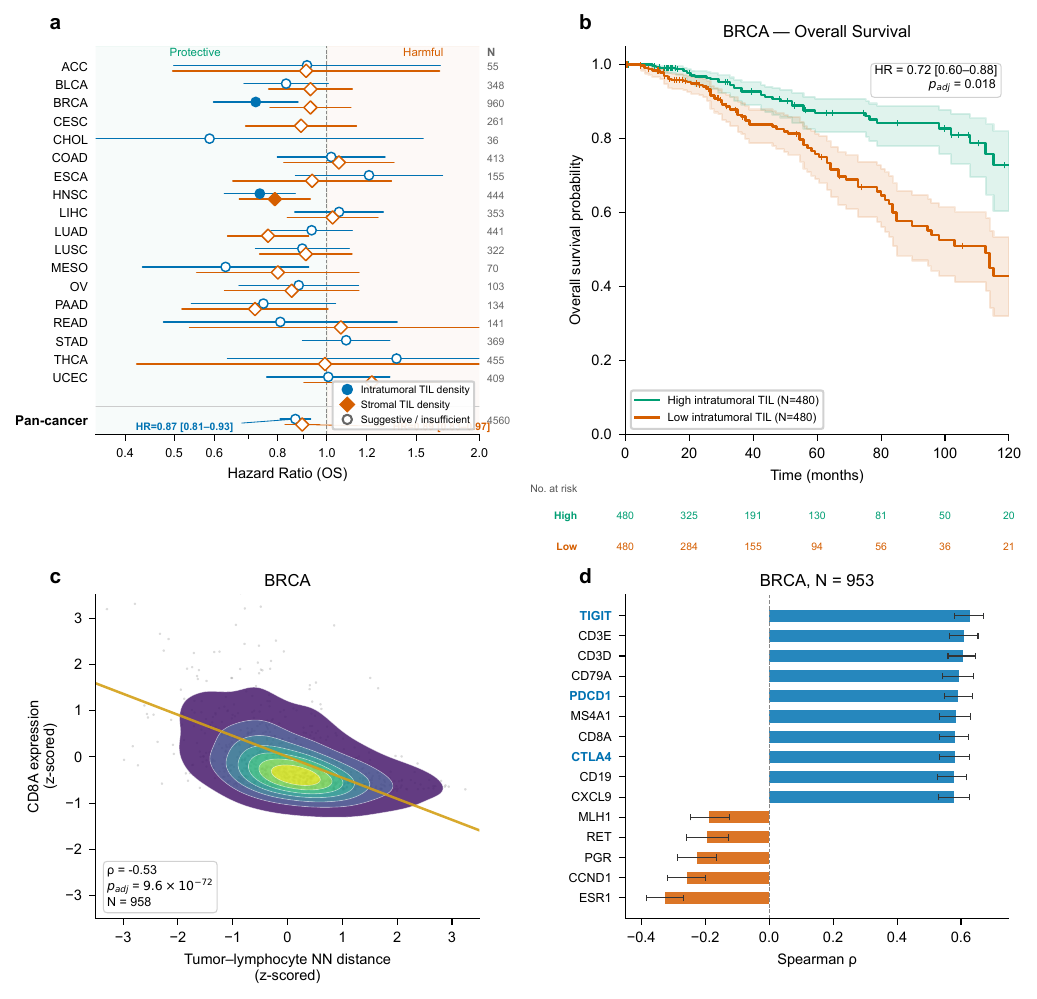}
  \caption{%
    \textbf{Spatial immune topology reveals compartment-specific prognostic effects.}
    (a)~Forest plot of hazard ratios (overall survival, covariate-adjusted Cox regression [age, sex, stage; stratified by TSS]) for intratumoral lymphocyte density (blue circles) and stromal lymphocyte density (orange diamonds) across cancer types and the pan-cancer cohort ($N = \num{4560}$). Filled markers indicate moderate or strong evidence (BH-adjusted $P < 0.05$ with adequate power); hollow markers indicate suggestive or insufficient evidence. Intratumoral lymphocyte density is protective (HR $= 0.87$ $[0.81, 0.93]$, $P_{\mathrm{adj}} = 9.8 \times 10^{-4}$); stromal lymphocyte density shows a weaker protective effect (HR $= 0.89$ $[0.83, 0.97]$, $P_{\mathrm{adj}} = 0.031$). Error bars represent 95\% confidence intervals. Vertical dashed line indicates HR $= 1.0$ (null).
    (b)~Kaplan--Meier curves for intratumoral lymphocyte density in BRCA (median split, $N = 960$; High: 480, Low: 480), showing a protective association (HR $= 0.72$ $[0.60, 0.88]$, $P_{\mathrm{adj}} = 0.018$). Shaded areas indicate 95\% confidence intervals. Number at risk shown below.
    (c)~Tumor--lymphocyte nearest-neighbor distance at the invasive front inversely correlates with CD8A expression in BRCA (Spearman $\rho = -0.53$, $P_{\mathrm{adj}} = 1.8 \times 10^{-68}$, $N = 958$), demonstrating that spatial immune exclusion detected by histomics corresponds to reduced cytotoxic T-cell gene expression. Per-slide feature values averaged per case; both axes z-scored within BRCA.
    (d)~Top gene correlates of intratumoral lymphocyte density in BRCA ($N = 953$, adjusted model). Horizontal bar chart showing the top 10 positive and top 5 negative Spearman correlations among significantly associated genes (BH-adjusted $P < 0.05$). Immune checkpoint genes (TIGIT, PDCD1, CTLA4) and T-cell markers (CD3E, CD3D, CD8A, CD8B) dominate the positive correlates, validating the biological identity of the histomic feature. Error bars represent 95\% bootstrap confidence intervals.
    All $P$-values were calculated using Cox proportional hazards regression (a,~b) or Spearman correlation with analytical $t$-test (c,~d), with Benjamini--Hochberg correction for multiple testing within each cancer type.
  }
  \label{fig:fig2}
\end{figure*}

\begin{figure*}[p]
  \centering
  \includegraphics[width=\textwidth]{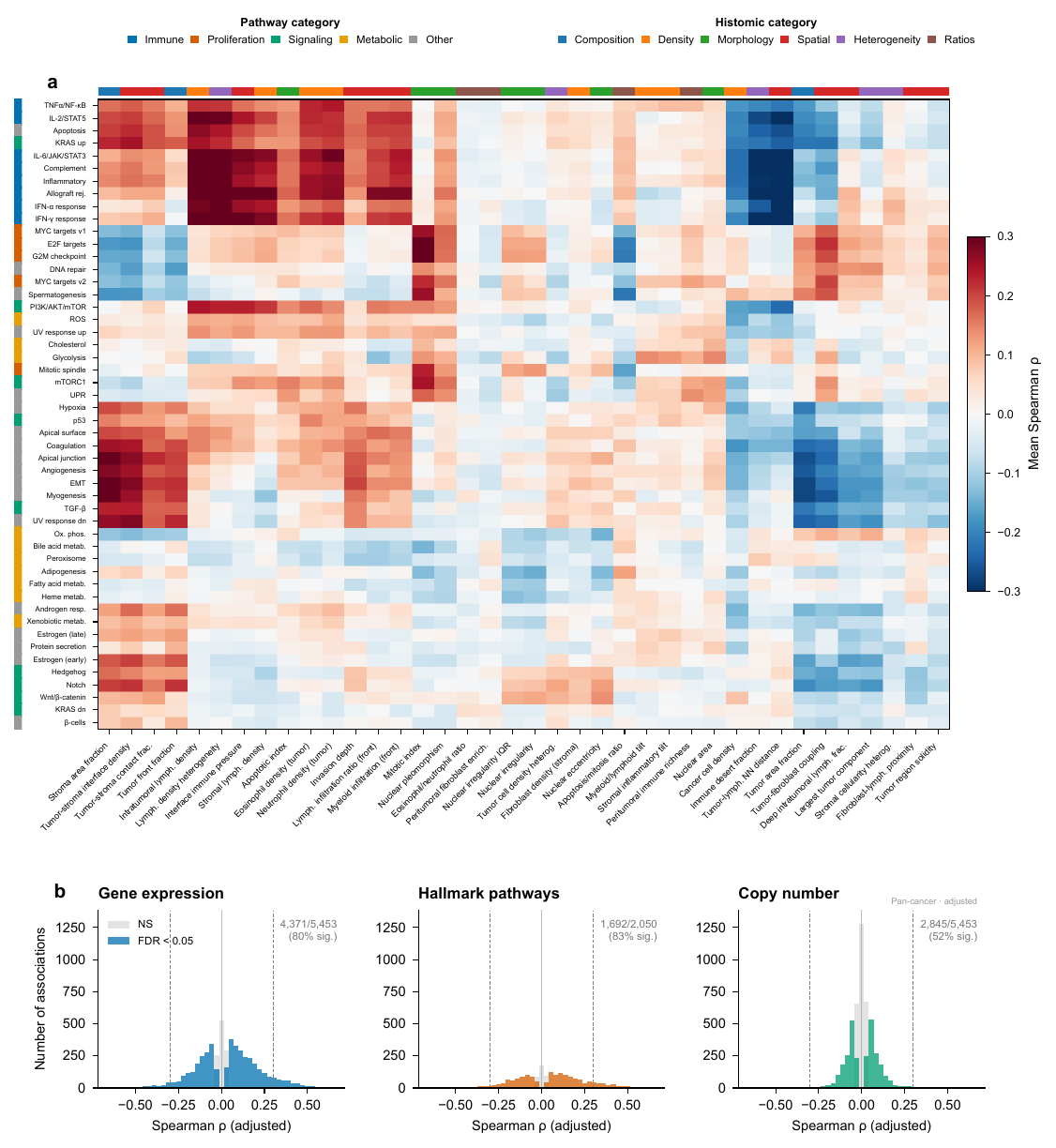}
  \caption{%
    \textbf{Morphological features recapitulate molecular programs.}
    (a)~Heatmap of mean Spearman correlation (across 21 cancer types) between 38 histomic features and 50 Hallmark pathway scores (unadjusted model). Rows and columns are hierarchically clustered (Ward linkage). Left color bar indicates pathway category (Immune, Proliferation, Signaling, Metabolic, Other); top color bar indicates histomic feature category (Composition, Density, Morphology, Spatial, Heterogeneity, Ratios). Structured correspondence is evident: immune cell density features cluster with immune pathway signatures; nuclear morphology and mitotic features cluster with cell cycle and proliferation pathways; invasion depth aligns with EMT. Colormap: RdBu\_r, clipped at $\rho = \pm 0.3$.
    (b)~Effect-size distributions for pan-cancer adjusted-model associations, stratified by molecular data type. Each histogram shows the distribution of Spearman $\rho$ values; colored bars indicate significance at FDR~$< 0.05$, gray bars indicate non-significant associations. Vertical dashed lines at $\rho = \pm 0.3$. Gene expression: \num{4371}/\num{5453} significant (80\%); Hallmark pathways: \num{1692}/\num{2050} (83\%); copy-number variation: \num{2845}/\num{5453} (52\%). The higher significance rate among pathway and expression associations, and the broader $\rho$ distributions, reflect stronger morphology--transcriptomic coupling than morphology--genomic coupling.
    All correlations are Spearman with analytical $P$-values ($t$-distribution approximation) and Benjamini--Hochberg correction.
  }
  \label{fig:fig3}
\end{figure*}

\begin{figure*}[p]
  \centering
  \includegraphics[width=\textwidth]{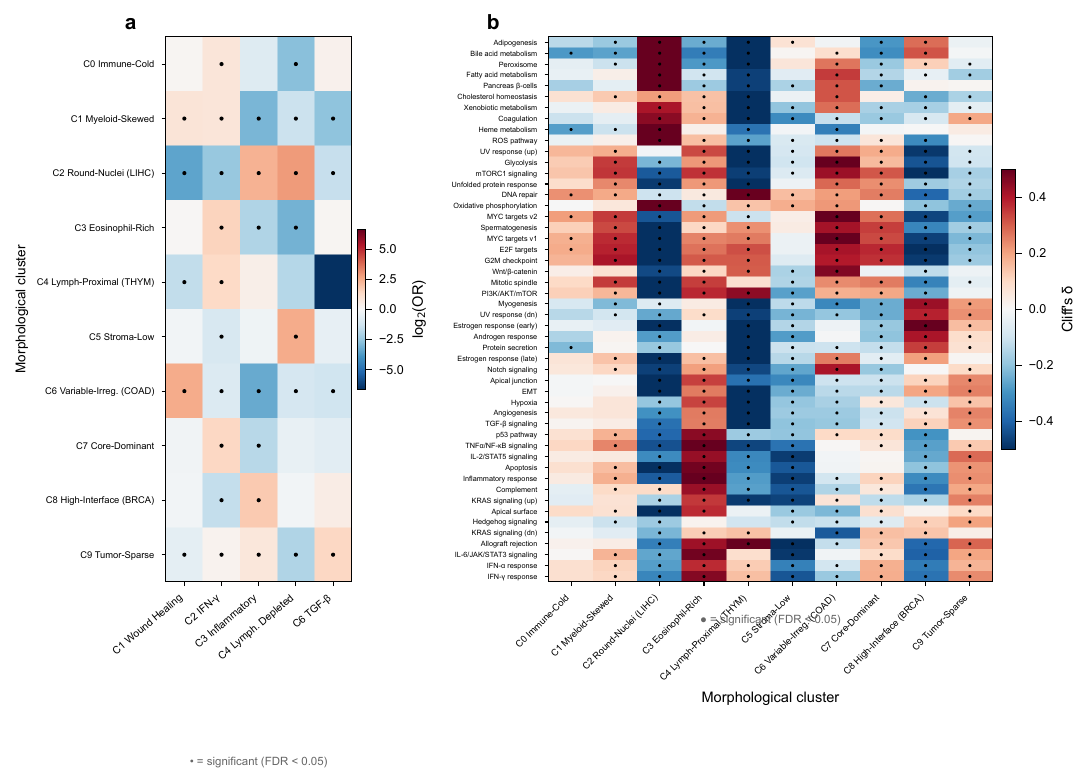}
  \caption{%
    \textbf{Morphological clusters map to distinct molecular archetypes.}
    (a)~Immune subtype enrichment per morphological cluster (L1, pan-cancer). Heatmap of $\log_2(\mathrm{OR})$ from Fisher's exact tests comparing the proportion of each Thorsson immune subtype within each cluster to the remaining cohort. Rows: 10 morphological clusters (labeled with cluster name and dominant cancer type). Columns: five Thorsson immune subtypes (C1 Wound Healing, C2 IFN-$\gamma$ Dominant, C3 Inflammatory, C4 Lymphocyte Depleted, C6 TGF-$\beta$ Dominant). Color scale: red--blue diverging, centered at 0. Black dots indicate BH-adjusted $P < 0.05$. Cluster~6 (CRC-enriched) is dominated by C1 Wound Healing (OR~$= 5.59$, $P_{\mathrm{adj}} < 10^{-88}$); Cluster~2 shows combined C4 Lymphocyte Depleted (OR~$= 7.14$) and C3 Inflammatory (OR~$= 4.99$) enrichment (85\% combined); Cluster~8 (hormone-driven) is enriched for C3 Inflammatory.
    (b)~Hallmark pathway enrichment (Cliff's $\delta$ from Mann--Whitney $U$ tests) per morphological cluster. Rows: 50 Hallmark pathways, hierarchically clustered (Ward linkage). Columns: 10 morphological clusters. Black dots indicate BH-adjusted $P < 0.05$. Cluster~4 (THYM-enriched) shows strong immune rejection pathway enrichment; Cluster~8 shows estrogen response enrichment ($\delta = 0.52$) with suppressed proliferation ($\delta = -0.51$); Cluster~6 shows Wnt/$\beta$-catenin enrichment ($\delta = 0.46$) consistent with CRC composition. Colormap centered at 0, range $[-0.5, 0.5]$.
  }
  \label{fig:fig4}
\end{figure*}

\begin{figure*}[p]
  \centering
  \includegraphics[width=\textwidth]{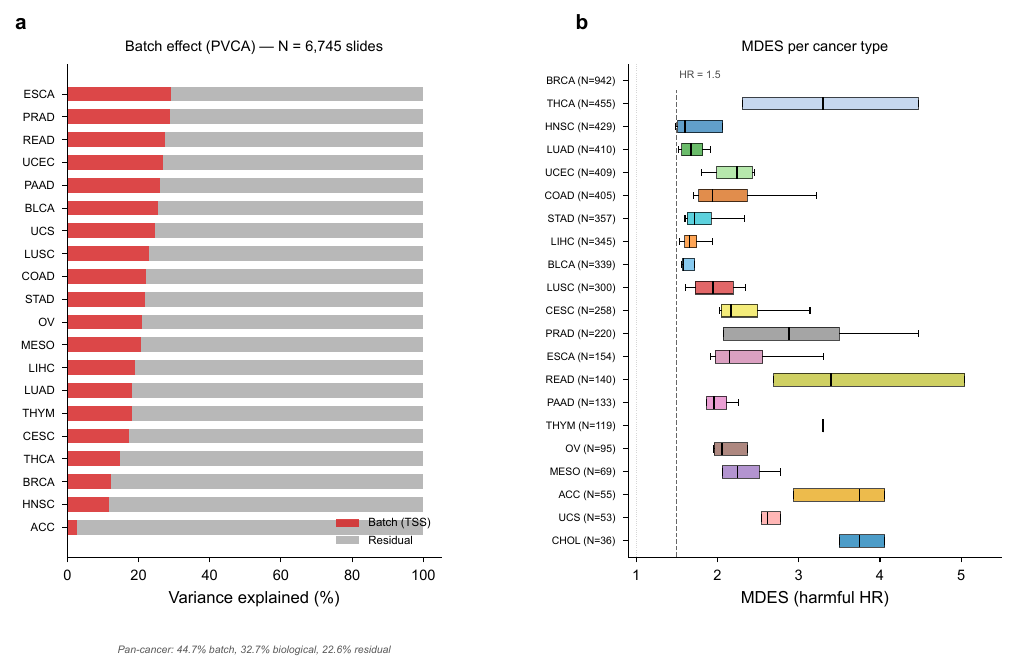}
  \caption{%
    \textbf{Statistical framework and quality control.}
    (a)~PVCA variance decomposition per cancer type showing proportions of variance attributable to batch effects (tissue source site, red) and residual signal (gray). Pan-cancer analysis attributes 44.7\% of variance to batch (TSS), 32.7\% to biological signal (cancer type), and 22.6\% to residual. Within individual cancer types, batch effects account for a median of 20.6\% of variance. All per-cancer silhouette scores by TSS are negative (range $-0.18$ to $-0.0004$), confirming minimal batch-driven clustering.
    (b)~Minimum detectable effect size (MDES) for harmful hazard ratios across 21 cancer types, ordered by sample size (ascending from bottom). Box plots show the distribution of MDES across features within each cancer type. Well-powered cancer types (BRCA, $N = 960$) can detect HR~$\geq 1.62$; underpowered types (CHOL, $N = 36$) require HR~$\geq 3.75$ for 80\% power. Dashed line indicates the clinically meaningful threshold (HR~$= 1.5$).
    $N$ values per panel: (a)~\num{6745} slides, 21 cancer types; (b)~\num{5623} survival associations across 22 cohorts.
  }
  \label{fig:fig5}
\end{figure*}

\begin{figure*}[p]
  \centering
  \includegraphics[width=\textwidth]{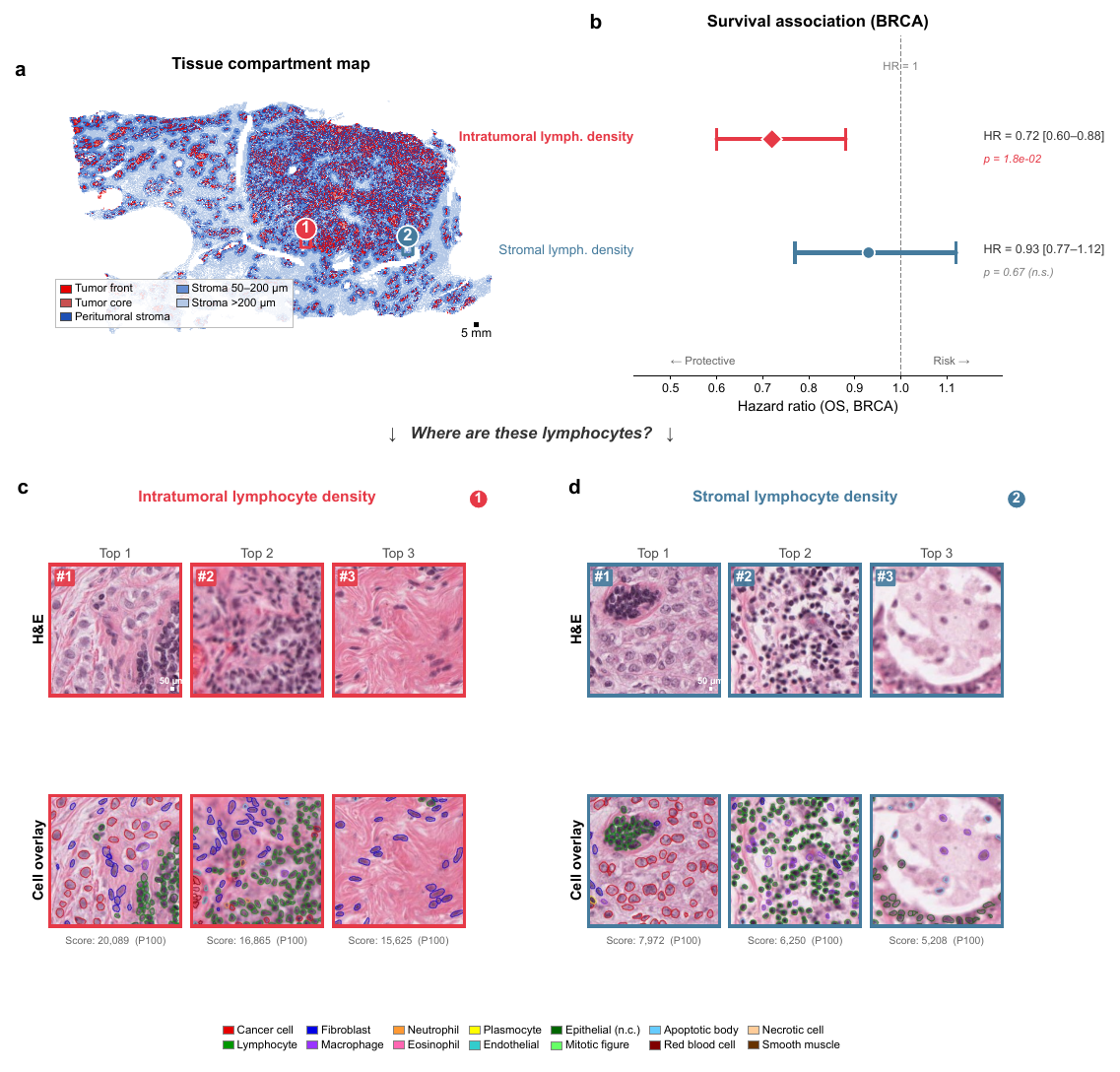}
  \caption{%
    \textbf{From statistics to cells: spatial interpretability in HistoAtlas.}
    (a)~Tissue compartment map for a representative BRCA slide (TCGA-A1-A0SE), showing a nine-class segmentation overlay: tumor front (red), tumor core (dark red), peritumoral stroma at three distance bands (0--50~\textmu m, 50--200~\textmu m, $>$200~\textmu m; blue shades), necrosis ring (brown), necrosis (gray), normal epithelium (green), and background (white). Numbered circles mark the tile regions shown in panels (c) and (d). Scale bar: 5~mm.
    (b)~Survival association (covariate-adjusted Cox regression, overall survival) for intratumoral and stromal lymphocyte density in BRCA ($N = 960$). Intratumoral lymphocyte density is protective (HR~$= 0.72$, 95\%~CI $[0.60, 0.88]$, $P_{\mathrm{adj}} = 0.018$); stromal lymphocyte density is not significant (HR~$= 0.93$, 95\%~CI $[0.77, 1.12]$, $P_{\mathrm{adj}} = 0.67$).
    (c)~Top-3 ranked tissue tiles for intratumoral lymphocyte density from the same slide, shown as H\&E (top) and cell-type prediction overlay (bottom) pairs. Dense green (lymphocyte) annotations among red (cancer cell) annotations confirm high intratumoral immune infiltration. Scale bar: 50~\textmu m.
    (d)~Top-3 ranked tiles for stromal lymphocyte density. Sparser lymphocyte annotations in stromal tissue visually mirror the weaker statistical association. Cell-type overlay legend (14 cell types) is shared between panels~(c) and~(d).
  }
  \label{fig:fig6}
\end{figure*}


\section{Discussion}
\label{sec:discussion}

HistoAtlas demonstrates that interpretable, spatially resolved histomic features
extracted from routine H\&E slides recapitulate canonical molecular
programs, including proliferation kinase networks, EMT transcriptional axes,
and immune cell gene signatures (Fig.~\ref{fig:fig3}), while stratifying
clinical outcomes across cancer types. The central advance is not any single
association but the comprehensive, statistically transparent linking of
\num{38} quantitative features to survival, gene expression, mutations, and
immune subtypes at pan-cancer scale. We deliver this linking as an openly
queryable resource.

A systematic biological plausibility audit
(Supplementary Table~\ref{tab:plausibility}) decomposed atlas findings into
60 atomic claims and assessed each against the literature. Of these,
42 (70\%) are well-established or supported by prior
studies~\cite{fridman2012immune, loi2019, denkert2018, nigg2001, nieto2016emt,
galon2006, pages2018, carretero2015, sahai2020, hatzikirou2012, giese2003glioma},
12 (20\%) are novel but biologically plausible, 5 (8\%) are novel with
uncertain mechanisms, and 1 (2\%) is an apparent contradiction that was resolved
upon examination: spatial composition heterogeneity does not equate to genetic
clonal diversity~\cite{marusyk2012}
(Supplementary Table~\ref{tab:plausibility}). No claim contradicted
established biology, a necessary consistency check for the feature extraction
and statistical framework.

Several atlas-enabled findings warrant targeted follow-up. The
compartment-specific difference in prognostic strength between intratumoral
and stromal lymphocyte density (strong versus weak protection,
respectively; Fig.~\ref{fig:fig2}a) is consistent with the importance of
immune cell localization within the tumor
microenvironment~\cite{galon2019immunoscore, chen2017oncology, joyce2015exclusion, fridman2012immune}.
Saltz et al.\ scored bulk TIL density across 13 TCGA cancer types from
deep-learning maps~\cite{saltz2018til} but did not distinguish intratumoral
from stromal compartments. To our knowledge, the differential prognostic
contribution of these compartments had not been quantified across 21 cancer
types from H\&E morphometrics.
We also identified morphologically distinct clusters with divergent survival
outcomes: quiescent-cold (Cluster~2; hazard ratio~$= 0.54$) versus hormone-driven (Cluster~8; hazard ratio~$= 1.37$; Fig.~\ref{fig:fig1}d). This finding suggests that
the binary immune-hot/immune-cold
classification~\cite{galon2019immunoscore} may obscure biologically and
potentially clinically relevant heterogeneity. Additional novel features showed consistent prognostic signals:
lymphocyte density spatial heterogeneity (coefficient of variation across
tiles) was protective in 14 of 17 evaluable cancer types (unadjusted model)
and may proxy tertiary lymphoid structure
formation~\cite{sautesfridman2019tls}; interface-normalized immune pressure,
a composite measure of immune cell engagement at the tumor-stroma boundary,
was protective in HNSC (hazard ratio~$= 0.74$,
$P_{\mathrm{adj}} = 3.9 \times 10^{-3}$); the
unadjusted associations previously observed in BRCA and LIHC did not survive
covariate adjustment. TCGA lacks immunotherapy response data,
so the clinical relevance of these immune distinctions for treatment
selection remains speculative. All findings are hypothesis-generating; none
should be interpreted as established biomarkers without independent
confirmation.

Evaluating these findings in context requires comparing HistoAtlas to existing
cancer data resources. cBioPortal~\cite{cerami2012cbio} provides comprehensive
molecular and clinical data but lacks any morphological features. The Human
Protein Atlas~\cite{uhlen2015hpa} provides semi-quantitative protein expression
scores from immunohistochemistry with cancer-specific survival
associations~\cite{uhlen2017pathology}, but does not extract continuous
morphometric features from H\&E-stained sections or link them to molecular
programs beyond single-protein correlations. The Cancer Imaging
Archive~\cite{clark2013tcia} hosts raw imaging data without a statistical
layer. Individual computational pathology studies have linked H\&E features
to outcomes in single cancer types~\cite{beck2011, yu2016}, and Diao et al.\
extracted 607 human-interpretable features across five cancer
types~\cite{diao2021hif}, but most recent approaches rely
on deep-learning embeddings that do not decompose into named histological
features~\cite{chen2024uni, vorontsov2024virchow}. HistoAtlas addresses this
interpretability gap: a tissue segmentation overlay with nine spatial zones derived from five tissue compartments, and cell-type annotations for nine morphological cell types enable users to trace any
statistical finding to specific tissue regions and verify the underlying cell
predictions visually (Fig.\,\ref{fig:fig6}a--d).

Three categories of limitation constrain the current atlas.
\emph{Data scope:} all \num{6745} slides derive from TCGA, a retrospective
convenience cohort with institutional selection biases~\cite{liu2018tcgacdr}.
We include 21 of 33 available cancer types; the 12 excluded types harbor
dominant cell populations (lymphoid, glial, melanocytic, mesenchymal,
neuroendocrine, renal tubular, or germ cell) outside the segmentation
model's training domain (Supplementary Table~3). TCGA participants are
predominantly of European
ancestry~\cite{liu2018tcgacdr, hoadley2018};
generalizability to diverse populations remains untested. Treatment standards
have evolved since TCGA accrual (2000--2016), limiting applicability to
contemporary regimens.
\emph{Feature quality:} because 13 of 21 cancer types are out-of-distribution for the cell segmentation model (trained on eight cancer types; \S\ref{sec:methods:features}), feature reliability varies across the atlas, and certain cancers (PRAD, LIHC, THCA) show
elevated mitotic and apoptotic indices from reduced cell detection. Seven ratio features required winsorization to
mitigate gate loophole artifacts, distance features are quantized at
\num{8}~\textmu m\,px$^{-1}$ resolution, and two tissue-model features carry
zero signal. We use one slide per case, sacrificing assessment of intratumoral
heterogeneity.
\emph{Analytical constraints:} all survival models test one histomic feature
at a time alongside clinical covariates; a penalized multivariate model
(e.g., LASSO Cox) would identify which features carry independent prognostic
information and is a natural next step. Several features are correlated by
construction (e.g., density features sharing the same denominator region),
so the effective number of independent features is lower than 38.
Morphology-to-molecular correlations are modest in magnitude (median
significant $|\rho| = 0.18$, IQR $0.13$--$0.27$); the features serve as
noisy proxies for, not replacements of, molecular measurements.
\emph{Validation gap:} we have not performed external replication. All
associations are internal to TCGA, and independent confirmation in
CPTAC~\cite{li2023cptac} or METABRIC~\cite{curtis2012metabric} cohorts is
required before any clinical interpretation. We have not benchmarked
interpretable features against whole-slide foundation model
embeddings~\cite{chen2024uni, vorontsov2024virchow}; a direct comparison of
predictive power versus interpretability would strengthen the case for
handcrafted features but requires a dedicated study.

Three extensions would substantially strengthen the atlas. Overlaying spatial
transcriptomics data (Visium, MERFISH) onto histomic features would provide
gold-standard validation for spatial immune metrics and could calibrate
morphometric proxies against measured transcript distributions. Integrating
foundation model embeddings alongside interpretable histomic features would
enable direct comparison of interpretability versus predictive power.
Extending the framework to non-TCGA cohorts would test generalizability and
enable community-contributed cancer types.

We designed HistoAtlas for transparency and reuse. Every association carries an
evidence-strength badge (strong, moderate, suggestive, or insufficient)
computed from adjusted $P$ values, effect sizes,
confidence interval widths, and sample sizes. The atlas reports the minimum
detectable effect size at 80\% power, conveying not only what it finds but
also what it cannot detect. Bidirectional spatial traceability links every
population-level statistic to tissue compartment maps and individual cell
annotations on the original slide, and from any slide back to
population-level associations (Fig.\,\ref{fig:fig6}c,d). By making every
morphological association traceable, statistically calibrated, and openly
queryable by humans and machines alike, HistoAtlas provides infrastructure for
systematic morphology-aware cancer analyses. All analysis code, feature
metadata, and precomputed results are publicly released.


\section{Methods}
\label{sec:methods}

\subsection{Data acquisition}
\label{sec:methods:data}

We obtained formalin-fixed, paraffin-embedded (FFPE) hematoxylin and eosin
(H\&E)-stained diagnostic whole-slide images from The Cancer Genome Atlas
(TCGA) via the Genomic Data Commons (GDC) portal for 21 solid-tumor cancer
types (Supplementary Table~3). Slides were excluded if the
viable tissue area fell below 1~mm$^2$, if severe processing artifacts (pen
marks covering $>$20\% of tissue area, out-of-focus regions) were present, or
if essential clinical metadata (vital status, follow-up time) was missing. To
avoid pseudo-replication, we retained one slide per case: for each case with
multiple diagnostic slides, we selected the primary tumor diagnostic slide with
the largest tissue area, yielding \num{6745} slides across \num{6745} unique
patients. Twelve additional TCGA cancer types were excluded because their
dominant cell morphologies fall outside the training domain of the cell
detection model (Supplementary Table~3).

Matched clinical data (overall survival, disease-specific survival,
disease-free survival, and progression-free survival; age at diagnosis, sex,
pathologic stage, tissue source site [TSS], and tumor purity estimates) were
obtained from the TCGA Pan-Cancer Clinical Data Resource
(TCGA-CDR)~\cite{liu2018tcgacdr}. Molecular data included RNA-seq gene
expression (RSEM normalized), somatic mutations from the MC3 multi-caller
ensemble~\cite{ellrott2018mc3}, copy-number variation, immune cell fraction
estimates from CIBERSORT~\cite{newman2015cibersort} and
xCell~\cite{aran2017xcell}, tumor purity from
ABSOLUTE~\cite{carter2012absolute}, and immune subtype classifications
(C1--C6) from Thorsson et~al.~\cite{thorsson2018immune}. All molecular data
were retrieved from the GDC and PanCancerAtlas data repositories. Molecular data were matched to slides by TCGA case barcode (first 12 characters of the barcode); for cases with multiple aliquots, the primary tumor aliquot was selected. Thorsson immune subtype labels (C1--C6) were matched to slides by TCGA case barcode. Of \num{6745} slides, \num{5590} (82.9\%) had matched immune subtype data; C5 (immunologically quiet, $n = 65$) and C6 (TGF-$\beta$ dominant, $n = 27$) were retained but had limited statistical power.

\subsection{Feature extraction}
\label{sec:methods:features}

We computed 38 quantitative histological-morphometric (hereafter ``histomic'')
features per slide, organized into five categories: tissue composition
(3~features), cell densities (6), nuclear morphology and kinetics (8), spatial
organization (18), and spatial heterogeneity (3). Complete definitions, units,
and category assignments are provided in
Supplementary Table~\ref{tab:feature_definitions}.

Feature extraction used two segmentation stages. Tissue segmentation used a
CellViT-inspired architecture~\cite{hoerst2024cellvit} with a Phikon
self-supervised ViT-B backbone~\cite{filiot2023phikon}, trained on the PanopTILs
crowdsourced annotation dataset~\cite{amgad2024panoptils}; model weights are
available in the code repository. We performed
inference at 0.5~\textmu m/px on $224 \times 224$~pixel tiles with 32-pixel
overlap; the final segmentation mask was obtained by majority voting in overlap
regions. The model classified each tile into nine tissue classes: cancerous
epithelium, stroma, necrosis, normal epithelium, TILs, junk/debris, blood,
other, and whitespace (mean intersection-over-union = 0.72 on the PanopTILs
held-out test set; note that per-class IoU varied substantially, and necrosis
and normal epithelium had near-zero recall in deployment, reflecting their
rarity in the training set and in resected TCGA specimens). Following the
International Immuno-Oncology Biomarker
Working Group recommendation~\cite{salgado2015til}, regions classified as TILs
were reclassified as stroma before all downstream computation, yielding five
effective compartments: cancerous epithelium, stroma, necrosis, normal
epithelium, and blood.

Cell segmentation and classification used the HistoPLUS
model~\cite{adjadj2025histoplus}, which detects and classifies individual cells
into nine morphological types: tumor cells, lymphocytes, fibroblasts,
plasmocytes, neutrophils, eosinophils, red blood cells, apoptotic bodies, and
mitotic figures (mean panoptic quality [PQ] = 0.509 across cell types; per-class
PQ varied from 0.28 to 0.73, with lowest performance on rare cell types such as
eosinophils and apoptotic bodies). Inference
was performed on $224 \times 224$~pixel tiles at 40$\times$ magnification
(0.25~\textmu m/px) when available, falling back to 20$\times$
(0.50~\textmu m/px); the majority of slides were scanned at 40$\times$.
We extracted tiles with a 64-pixel (16~\textmu m)
overlap margin; cells detected in overlap regions were deduplicated via a
union-find algorithm that merges instances whose centroids fall within
10~\textmu m, corresponding approximately to the diameter of a typical
epithelial nucleus. The cell model was trained on pathologist annotations from
eight cancer types (LUAD, LUSC, BRCA, COAD, BLCA, OV, PAAD, MESO); the
remaining 13 included cancer types were processed in an out-of-distribution
(OOD) setting. Tissue--cell discordance (cells in a tissue-model tumor region
not classified as cancer cells by the cell model) varied from 0\% to 12\%
across all cancer types, with the highest rate in PAAD (12.2\%) despite being
in-distribution, likely due to the dense desmoplastic stroma and small tumor
glands characteristic of pancreatic ductal adenocarcinoma. OOD cancer types
showed additional issues, including inflated kinetic indices in PRAD, LIHC, and
THCA attributable to reduced cell-detection sensitivity. Red blood cells and plasmocytes were not used in
feature computation.

All spatial features were computed on compartment masks resampled to a common
resolution of $r = 8$~\textmu m/px by nearest-neighbor interpolation to ensure
scanner-invariant boundary computation. Connected components below
$A_{\min} = \num{2048}$~\textmu m$^2$ (${\approx}\,32$~pixels at
8~\textmu m/px, equivalent to approximately $5 \times 5$ cell diameters) were
removed per compartment before distance transform
computation to prevent noisy segmentation fragments from inflating boundary
lengths. Five spatial bands were defined using the signed Euclidean distance
transform $d_T$ from the tumor boundary, where the tumor boundary was defined
as the outer contour of the cancerous epithelium compartment mask at
8~\textmu m/px resolution (positive inside tumor, negative outside): tumor front $B_T^{0\text{-}50}$ ($0 \leq d_T \leq 50$~\textmu m),
tumor core $B_T^{>50}$ ($d_T > 50$~\textmu m), stroma near
$B_S^{0\text{-}50}$ ($-50 \leq d_T < 0$), stroma far $B_S^{50\text{-}200}$
($-200 \leq d_T < -50$), and necrosis ring $R_{\text{Nec}}^{0\text{-}100}$
(within 100~\textmu m of the necrosis boundary). The 50~\textmu m front band
corresponds to approximately five cell diameters; the 200~\textmu m stroma
cutoff was chosen as a heuristic approximation of the attenuation range of
immune infiltration gradients, informed by spatial immune profiling
studies~\cite{saltz2018til, keren2018mibi}. We classified slides into
growth-pattern regimes based on the tumor front fraction
$\varphi = A(B_T^{0\text{-}50}) / A(\Omega_T)$: mass-forming
($\varphi \leq 0.5$), intermediate ($0.5 < \varphi \leq 0.8$), or
infiltrative ($\varphi > 0.8$). A macro-tumor mask obtained by morphological
closing (disk radius $\rho = 200$~\textmu m), used solely for computing the
\texttt{micro\_interface\_ratio} QC metric, detected micro-interface
dominance in infiltrative tumors.

Two additional features (normal epithelium area fraction and tumor--normal
contact fraction, both dependent on normal epithelium detection, which the
tissue model did not reliably identify) were excluded from the atlas entirely
due to zero signal across all slides. Seven ratio features were susceptible to
gate-loophole artifacts producing extreme values when denominators approached
zero; these were mitigated by the
winsorization step described below.
Distance-based features (nearest-neighbor distances) were quantized at
8~\textmu m/px resolution owing to the tile grid spacing.

\subsection{Feature preprocessing}
\label{sec:methods:preprocessing}

Preprocessing of the 38 histomic features proceeded in three steps. First, 22
features with heavy right-skew (all cell densities, ratio features, distance
features, and heterogeneity measures including coefficients of variation) were
log-transformed using $\log(1 + x)$. Second, all 38 features were winsorized at
the 0.5th and 99.5th percentiles (computed per feature across the full cohort)
to mitigate the influence of extreme values arising from segmentation artifacts.
Global (pan-cohort) percentiles were used because the target
artifacts (segmentation failures and ratio instabilities) are present across
cancer types, though at varying rates (tissue--cell discordance ranges from 0\%
to 12\% by cancer type). Per-cancer-type winsorization was
not used because the small sample sizes of several cohorts (e.g., CHOL,
$n = 38$) would yield unreliable percentile estimates. Third, features were
$z$-score standardized to zero mean and unit variance.
The scope of standardization varied by analysis: UMAP embeddings and
$K$-means clustering used global $z$-scores (across all $n$ samples), Cox
regression used per-cancer-type $z$-scores (so hazard ratios reflect per-SD
effects within each cancer type), and Spearman correlations used no $z$-score
standardization (rank-based). Missing values ($<$2\% of entries across all
features) were imputed with the column
median prior to standardization. Raw (untransformed) values were retained for
display in the web atlas interface; all statistical models and embeddings
consumed preprocessed values.

\subsection{Survival analysis}
\label{sec:methods:survival}

We assessed associations between each histomic feature and four survival
endpoints (overall survival [OS; designated as the primary endpoint],
disease-specific survival [DSS], disease-free survival [DFS], and
progression-free survival [PFS]) using univariate Cox proportional
hazards (PH) regression, implemented with the \texttt{lifelines} Python
package~\cite{davidsonpilon2019} (no L2 penalization;
\texttt{penalizer=0.0}). Each model tests a single histomic feature (plus
covariates); multivariate models incorporating multiple features simultaneously
are not included in the current atlas (see Discussion). For each
feature--cancer type--endpoint
combination, we fitted two models: (1)~unadjusted (feature only); and
(2)~adjusted (feature + age at diagnosis + sex + pathologic stage as covariates,
stratified by tissue source site [TSS]). TSS was handled as a stratification
variable in Cox models (\texttt{strata=['tss']} in \texttt{lifelines}), allowing
each site to have its own baseline hazard function without consuming degrees of
freedom, the standard approach for multi-center
studies~\cite{therneau2000modeling}. Hazard ratios (HR) were reported per one
standard deviation increase in the preprocessed feature value, with 95\% Wald
confidence intervals computed as
$\exp(\hat{\beta}_1 \pm \Phi^{-1}(0.975) \cdot \mathrm{SE}(\hat{\beta}_1))$,
where SE was derived from the observed Fisher information matrix.
Pan-cancer analyses used stratified Cox regression with cancer type as an
additional stratification variable (\texttt{strata=['cancer\_type']} in
\texttt{lifelines}), allowing each cancer type its own baseline hazard function
while estimating a single shared regression coefficient. This approach avoids
confounding by differential baseline hazard rates across cancer types.
Per-cancer-type results constitute the primary analyses. Features and age were
standardized (z-scored) prior to fitting; categorical covariates (sex, stage)
were one-hot encoded with the first category dropped.
For the adjusted model, we used multiple imputation by chained equations (MICE)
with a Bayesian ridge regression imputer (\texttt{scikit-learn}
\texttt{IterativeImputer}, 5 imputations, max 10 iterations per imputation,
random seed incremented per imputation). Results were pooled using Rubin's
rules, yielding combined hazard ratios, standard errors, and $P$-values; the
fraction of missing information (FMI) was recorded for each analysis. When
covariate missingness exceeded 20\% in any variable, when total sample size was
below 30, or when fewer than 3 of 5 imputations produced valid fits, MICE was
skipped and complete-case analysis was used as a fallback. Missingness arose
primarily from incomplete staging annotations and varied by cancer type
(Supplementary Table~\ref{tab:complete_case_n}). A minimum sample size of
$n \geq 30$ with $\geq$10 events was required per feature--cancer
type--endpoint combination. Models were fitted only when the residual degrees of
freedom remained positive after accounting for all covariates.

We tested the proportional hazards assumption for every fitted model using
Schoenfeld residuals with the Kaplan--Meier time
transform~\cite{grambsch1994}. The PH test $P$-value was used to
flag each association: ``pass'' ($P \geq 0.05$), ``warn'' ($0.01 \leq P < 0.05$),
or ``fail'' ($P < 0.01$). The $P < 0.01$ threshold for failure was chosen to
balance sensitivity against the high multiple-testing burden (hundreds of models
per cancer type); we note that this threshold has asymmetric power: large
cohorts (e.g., BRCA, $n \approx 960$) can detect trivial PH departures, whereas
small cohorts (e.g., CHOL, $n = 36$) have limited power to detect even
substantial violations. The prevalence of PH violations across cancer types and
endpoints is reported in Supplementary Table~7. When the PH assumption was violated ($P < 0.05$, encompassing both ``warn''
[$0.01 \leq P < 0.05$] and ``fail'' [$P < 0.01$] tiers), we invalidated the Cox
hazard ratio, confidence intervals, and $P$-value (set to NaN), as these
quantities are unreliable under non-proportional hazards. For BH correction,
invalidated $P$-values were set to 1.0 to preserve the correction family size
without inflating false discoveries. As a complementary, assumption-free
summary, we additionally computed the restricted mean survival time (RMST)
difference.

RMST was calculated as the area under the Kaplan--Meier curve up to a
cancer-type-specific time horizon, with standard errors estimated using the
Irwin variance formula~\cite{irwin1949}. The default horizon was
\num{1095}~days (3~years). For aggressive cancer types with short median
survival, we used a 2-year horizon (PAAD, MESO). For indolent cancer types, we
used a 5-year horizon (THCA, PRAD). These horizons were chosen to ensure
adequate follow-up and at-risk populations at the truncation point;
cancer-type-specific values are listed in Supplementary
Table~\ref{tab:rmst_horizons}. RMST differences between high and low feature
groups (median split) were tested using a permutation procedure (\num{5000}
permutations of group labels). The median split was used for interpretability
and consistency with the clinical convention of risk stratification; we
acknowledge that dichotomization discards information and reduces statistical
power relative to continuous methods~\cite{royston2006}. $P$-values were
computed as
$P = (\#\{b : |\Delta_b| \geq |\Delta^{\mathrm{obs}}|\} + 1) / (B + 1)$.
Bootstrap confidence intervals (95\%) for the RMST
difference were obtained from \num{1000} bootstrap resamples of the full cohort,
using the percentile method. RMST analyses required $n \geq 20$ total subjects
with $\geq$5 observations per group.

For cluster-level survival comparisons, a two-sided log-rank test was computed
between cluster members and non-members, with BH correction applied within each
combination of cluster level, analysis type, cancer type, and endpoint.

\subsection{Molecular correlations}
\label{sec:methods:correlations}

We computed Spearman rank correlations between each of the 38 histomic features
and 293 molecular targets: 133
curated genes drawn from established cancer gene panels (OncoKB~\cite{oncokb},\allowbreak{} COSMIC Cancer Gene Census~\cite{sondka2018cgc}),\allowbreak{} immune checkpoint targets (TIGIT, PD-L1, CTLA4, LAG3),\allowbreak{} and EMT/stemness markers curated from Nieto et~al.~\cite{nieto2016emt} (complete list in Supplementary Table~4), each assessed for
both expression and copy-number variation (133 $\times$ 2 = 266); 21 MSigDB
Hallmark pathway scores~\cite{liberzon2015} (of the 50 Hallmark gene sets [Supplementary Table~\ref{tab:pathway_signatures}], 21 had sufficient matched data after intersection filtering); and 6 immune cell fraction
scores from CIBERSORT~\cite{newman2015cibersort}. Pathway activity scores
were computed via single-sample Gene Set Enrichment Analysis
(ssGSEA; \texttt{gseapy}~1.1.12)~\cite{barbie2009ssgsea} using the full TCGA Pan-Cancer batch-corrected
RNA-seq expression matrix (${\sim}$20,500 genes per sample,
EB++Adjust\-PANCAN\_\allowbreak{}Illumina\-HiSeq\_\allowbreak{}RNASeqV2 from UCSC Xena~\cite{goldman2020xena}).
The 50 Hallmark gene sets (median ${\sim}$200 genes per pathway) provide
well-characterized, non-redundant representations of biological processes
with established provenance~\cite{liberzon2015}. Gene sets with
fewer than 10 genes present in the expression matrix after intersection
were excluded (a stricter threshold than GSEA enrichment analysis below, because per-sample ssGSEA scores are noisier with small gene sets).

For unadjusted models, we computed the standard Spearman correlation
coefficient and its analytical $P$-value ($t$-approximation) using
\texttt{scipy}\allowbreak\texttt{.stats}\allowbreak\texttt{.spearmanr}, which is accurate for $n \geq 10$ and avoids
the $P$-value banding that occurs with permutation floors. Bootstrap confidence
intervals (95\%) were obtained from \num{1000} resamples using the percentile
method: for each resample, observations were drawn with replacement, ranks
recomputed, and the Spearman $\rho$ recorded.

For covariate-adjusted models, we computed partial Spearman correlations via the
following procedure: (1)~rank-transform both the histomic feature and the
molecular target, as well as all covariates; (2)~residualize the ranked feature
and ranked target against the ranked covariates using ordinary least-squares
regression; and (3)~compute the Pearson correlation of the residuals, yielding
the partial Spearman $\rho$~\cite{conover1999}. $P$-values for
partial correlations were obtained using the standard $t$-test for partial
correlations: $t = r\sqrt{{\mathrm{df}} / ({1 - r^2})}$ with
$\mathrm{df} = n - 2 - k$, where $k$ is the number of covariate columns after
one-hot encoding, and a two-sided $P$-value from the $t(\mathrm{df})$
distribution. This matches the approach used by R's
\texttt{ppcor::pcor.test()} and Python's
\texttt{pingouin.partial\_corr()}. Bootstrap CIs
for the partial Spearman $\rho$ were obtained from \num{1000} resamples, each
recomputing the full rank--residualize--correlate pipeline.

A cancer type was included in the correlation analysis if it contained $n \geq 30$
samples; individual feature pairs required $n \geq 10$ non-missing observations.
All correlations were computed per cancer type. Deterministic per-task random seeds
were derived from a hash of the cancer type, histomic feature, and molecular
feature names, ensuring reproducibility across parallel executions.

In total, the correlation analysis comprised \num{487638} histomic--molecular
pairs: 38 features $\times$ 293 molecular targets (133 genes $\times$ 2 data
types [expression and copy number] $+$ 21 Hallmark pathway scores $+$ 6 immune cell
scores) $\times$ 22 cohorts (21 cancer types $+$ pan-cancer), each evaluated
under two adjustment models (unadjusted and adjusted [age, sex, stage, TSS]);
the stated total excludes combinations with insufficient data. TSS was included
as a grouped covariate rather than a stratification variable because the
rank-residualization procedure requires explicit covariate columns; TSS was
grouped into the five most frequent sites per cancer type, with remaining sites
collapsed to ``Other'', to limit the number of dummy variables in the
residualization.

\subsection{Categorical associations}
\label{sec:methods:categorical}

Associations between histomic features and categorical molecular variables
(somatic mutations [mutant vs.\ wild-type], copy-number alterations
[amplification/deletion vs.\ neutral], and immune subtypes [C1--C6]) were
tested separately for each cancer type.

For unadjusted models, inference used the exact or asymptotic distribution
of the test statistic as implemented in \texttt{scipy}\allowbreak\texttt{.stats}. Two-group
comparisons used the Mann--Whitney $U$ test (two-sided) with Cliff's $\delta$ as
the effect size. Cliff's $\delta$ was
computed as the mean of the sign matrix $\text{sign}(x_i - y_j)$ over all
$n_1 \times n_2$ pairs, with 95\% bootstrap CIs (\num{1000} resamples);
bootstrap rather than analytical CIs were used because the data contain tied
values, which violate the continuity assumption of
Cliff's analytical variance formula~\cite{cliff1993}. For
variables with more than two levels, we used the Kruskal--Wallis $H$ test with
$\eta^2_H = (H - k + 1)/(N - k)$ as the effect
size~\cite{tomczak2014}, floored at zero. Bootstrap CIs for
$\eta^2$ were obtained from \num{1000} resamples of the group arrays.

For covariate-adjusted models, we used rank-ANCOVA with Freedman--Lane
permutation inference~\cite{freedmanlane1983}. The response (histomic feature) and all covariates were
rank-transformed. The group effect was tested by comparing the full model
(group dummies + rank-transformed covariates) to the reduced model (covariates
only) via an $F$-statistic on the residual sum of squares. $P$-values were
obtained by permuting residuals from the covariate-only model (\num{1000}
permutations). The $F$-statistic null distribution is parametrically smooth
(approximately $F$-distributed under the null), requiring relatively few
permutations for stable $P$-value estimation; with \num{1000} permutations
the minimum achievable $P$-value is ${\approx}0.001$, sufficient for BH
correction within the per-cancer-type families used here. The
effect size was partial $\eta^2 =
\text{SS}_{\text{group}} / (\text{SS}_{\text{group}} + \text{SS}_{\text{residual}})$.
Bootstrap CIs for partial $\eta^2$ were computed from \num{1000} resamples.

Group ordering for binary comparisons followed a deterministic convention:
mutant before wild-type, amplification/deletion before neutral. This ensured
consistent sign interpretation of Cliff's $\delta$ across analyses. A minimum of
$n \geq 30$ total observations and $\geq$5 observations per group was required;
groups smaller than 5 were excluded.

\subsection{Clustering}
\label{sec:methods:clustering}

Unsupervised morphological clustering was performed in two tiers. At the L1
(pan-cancer) level, all \num{6745} slides were clustered on the full
38-dimensional preprocessed feature vector (after log-transform, winsorization,
and $z$-scoring). At the L2 (cancer-specific) level, clustering was performed
independently within each cancer type using cancer-type-specific $z$-scores.

We used $K$-means clustering (\texttt{scikit-learn}, $n_{\text{init}} = 10$,
random state fixed at 42). For L1, clustering was computed for $K \in \{3, \ldots, 25\}$;
silhouette, Calinski--Harabasz, Davies--Bouldin, and gap statistic scores were computed for each $K$.
$K = 10$ was selected based on convergence of silhouette (local maximum at $K = 10$), Calinski--Harabasz (plateau), Davies--Bouldin (local minimum), and gap statistic scores, balancing cluster interpretability against granularity (the gap statistic favored $K = 8$; silhouette favored $K = 10$). $K$-means was chosen for its
scalability and interpretability; the resulting clusters should be understood as
a convenient partition of feature space rather than a claim about the true number
of distinct morphological subtypes. We did not apply PCA dimensionality reduction before clustering; the 38-dimensional feature space was used directly to preserve interpretability of cluster feature profiles. The effective dimensionality (17 components explain 90\% of variance) suggests moderate redundancy, which gives correlated features proportionally more weight. For L2,
the number of clusters was selected by the elbow method: we computed inertia
for $K$ values ranging from 2 to 8 (depending on cohort size) and selected the
$K$ that maximized the perpendicular distance from each point to the line
connecting the first and last $(K, \text{inertia})$ points in the normalized
space~\cite{thorndike1953}. Cancer types with fewer than 20 samples were assigned a single cluster without
optimization. For cancer types with $\geq$20 samples, this yielded
$K \in [2, 7]$ (69 L2 subclusters total). Individual clusters
with fewer than 10 samples were excluded. Internal validation metrics
(silhouette score, Calinski--Harabasz index, and Davies--Bouldin index) were
computed for each candidate $K$ and reported alongside the selected solution.

Cluster stability was assessed via repeated random subsampling: 50 iterations of 80\%
subsampling without replacement, re-clustering with the same $K$ and
$n_{\text{init}} = 10$, and comparison to the original labels using the adjusted
Rand index (ARI) and mean best-match Jaccard index across
clusters~\cite{hennig2007}. The 50-iteration count provides sufficient precision
for stability estimation: at ARI~$= 0.72$, the Monte Carlo standard error is
${\approx}0.02$, small relative to the typical gap between stable and unstable
solutions. Cluster names were automatically
generated from a structured schema. Each cluster received an immune axis label
(immune-hot, immune-cold, or immune-mixed, based on mean $z$-score $> 0.5$ or
$< -0.5$ across five immune features), a stromal axis label (stroma-high,
stroma-low, or stroma-mid), the label of the most extreme non-immune,
non-stromal feature, and a cancer-enrichment tag when a single cancer type
constituted $\geq$40\% of the cluster.

Two-dimensional visualization was performed using UMAP (uniform manifold
approximation and projection) with $n_{\text{neighbors}} = 15$,
$\text{min\_dist} = 0.1$, Euclidean metric, and random state 42, applied to
the $z$-scored feature matrix after median imputation of missing
values~\cite{mcinnes2018}. UMAP was used for visualization only; no statistical
inference was drawn from the embedding coordinates, and all clustering was
performed in the original 38-dimensional feature space, so UMAP hyperparameter
sensitivity does not affect statistical conclusions.

\subsection{Cluster enrichment}
\label{sec:methods:enrichment}

\paragraph{Mutation enrichment.}
Mutation enrichment for each cluster was tested using Fisher's exact test on
$2 \times 2$ contingency tables (mutated vs.\ wild-type $\times$ in-cluster
vs.\ out-of-cluster), with odds ratios (OR) and exact 95\% confidence
intervals. Fisher's exact test conditions on both margins and is therefore
conservative when only the row margin (cluster membership) is fixed; we accepted
this conservatism in exchange for exact $P$-values without distributional
assumptions. A minimum of 20 total observations and 5 mutated samples were
required per test.

\paragraph{Pathway enrichment.}
Pathway enrichment was assessed using two complementary approaches. First,
for each cluster, we compared the distribution of pathway activity scores
(in-cluster vs.\ out-of-cluster) using the Mann--Whitney $U$ test with
Cliff's $\delta$ as the effect size and \num{1000}-resample bootstrap CIs. Second,
we performed gene set enrichment analysis
(GSEA)~\cite{subramanian2005} with phenotype permutation. Gene-level
$t$-statistics were computed from Welch's $t$-test comparing in-cluster to
out-of-cluster expression. The enrichment score (ES) was calculated as the
weighted running sum statistic using absolute $t$-statistics as weights.
Significance was assessed by permuting sample phenotype labels (\num{1000}
permutations). $P$-values were computed separately for positive and negative
enrichment:
\[
P = \begin{cases}
(\#\{b : \mathrm{ES}_b^{+} \geq \mathrm{ES}^{\mathrm{obs}}\} + 1) /
  (\#\{\mathrm{ES}_b^{+}\} + 1) & \text{if } \mathrm{ES}^{\mathrm{obs}} > 0, \\[4pt]
(\#\{b : \mathrm{ES}_b^{-} \leq \mathrm{ES}^{\mathrm{obs}}\} + 1) /
  (\#\{\mathrm{ES}_b^{-}\} + 1) & \text{if } \mathrm{ES}^{\mathrm{obs}} < 0,
\end{cases}
\]
where $\mathrm{ES}_b^{+}$ and $\mathrm{ES}_b^{-}$ denote the positive and negative
tails of the null distribution, respectively.
Normalized enrichment scores (NES) were
obtained by dividing the observed ES by the mean of the positive or negative
tail of the null distribution. False discovery rates (FDR $q$-values) were
computed using the pooled null NES distribution across all gene sets, following
the standard GSEA procedure~\cite{subramanian2005}. Gene sets with
FDR $q < 0.25$ were considered significant. The FDR $q < 0.25$ threshold follows the original GSEA convention~\cite{subramanian2005}, reflecting the exploratory nature of pathway enrichment and the lower statistical power of rank-based enrichment relative to parametric tests used elsewhere. Leading-edge genes (up to 20 per
pathway) were recorded. Gene sets with fewer than 3 genes present in the
expression matrix after intersection were excluded.

\paragraph{Immune subtype enrichment.}
Immune subtype enrichment (Thorsson C1--C6 classification) was tested per
cluster using Fisher's exact test on $2 \times 2$ tables (subtype present/absent
$\times$ in-cluster/out-of-cluster), with odds ratios, exact 95\% CIs, and
observed-to-expected ratios.

\subsection{Multiple testing correction}
\label{sec:methods:multiple_testing}

All $P$-values were corrected for multiple comparisons using the
Benjamini--Hochberg (BH) procedure~\cite{benjamini1995} applied
within explicitly defined correction families. Each family was defined to
group biologically coherent tests while avoiding excessive conservatism from
pooling unrelated analyses (Supplementary Table~6 lists the exact family
definition for each analysis type). For example, survival correction families
were defined per cancer type $\times$ endpoint $\times$ adjustment model,
ensuring that BH correction pools only the 38 feature-level tests within
a single biological context. GSEA used its own
canonical FDR procedure based on pooled null NES distributions rather than BH
correction~\cite{subramanian2005}. Every adjusted
$P$-value was stored alongside the correction family identifier and the number
of tests in the family (\texttt{correction\_family\_id},
\texttt{n\_tests\_in\_family}) to enable post hoc verification.

OS was designated as the primary survival endpoint; DSS, DFS, and PFS are
reported as sensitivity analyses. BH correction was applied independently per
endpoint rather than pooling across endpoints, which is appropriate because the
endpoints have partially overlapping event definitions (OS and DSS share events)
and thus violate the independence assumption of joint correction.
Cross-endpoint replication (i.e., the fraction of OS-significant associations
that also reach significance for DSS or PFS) is reported as an informal
consistency check rather than a formal multiplicity-controlled comparison.

\paragraph{Statistical conventions.}
Permutation $P$-values (RMST, rank-ANCOVA) used the add-one
correction of Phipson and Smyth~\cite{phipson2010permutation}:
$P = (B^+ + 1) / (B + 1)$. Bootstrap confidence intervals (\num{1000}
resamples, percentile method) were computed for all effect sizes; CIs were
reported only when $\geq$90\% of resamples produced valid estimates (a resample
was invalid when zero-variance columns or singular covariate matrices prevented
model fitting). One thousand resamples provide adequate precision for central tendency estimation (Monte Carlo SE $< 0.01$ for moderate effect sizes); tail coverage may be imprecise for extreme quantiles. Permutation counts were set to \num{5000} for RMST (where
permutation is the sole inference method), \num{1000} for GSEA (with \num{1000}
permutations the minimum achievable $P$-value is ${\approx}0.001$, sufficient for
the BH correction applied within per-cluster
families), and \num{1000} for rank-ANCOVA (where the
$F$-statistic null distribution is approximately parametric, yielding stable
$P$-value estimates with fewer permutations).
Minimum sample sizes were set based on the number of parameters estimated: $n \geq 30$ for regression models with covariates (Cox, Spearman, rank-ANCOVA), $n \geq 20$ for nonparametric group comparisons (RMST, Fisher's exact test).
Ties were handled using midranking for Spearman correlations and the normal approximation with continuity correction for Mann--Whitney $U$ tests, as implemented in \texttt{scipy.stats}. Throughout the text, statistical significance
refers to $P_{\text{adj}} < 0.05$ (BH-corrected) unless otherwise stated.

\subsection{Batch effect assessment}
\label{sec:methods:batch}

We assessed potential batch effects from tissue source site (TSS) using two
complementary approaches. Principal variance component analysis
(PVCA)~\cite{li2009pvca} decomposed variance in the top principal
components (selected to explain $\geq$80\% cumulative variance, up to 10
components) into TSS (batch), cancer type (biological), and residual fractions
using marginal one-way ANOVA $\eta^2$ for each factor (TSS, cancer type) applied
independently to each principal component, weighted by explained variance
ratio. Because TSS and cancer type are not orthogonal (most sites contribute
predominantly one cancer type), the marginal $\eta^2$ values can sum to more
than 1.0 per component; residual variance was computed as
$\max(0, 1 - \eta^2_{\text{batch}} - \eta^2_{\text{bio}})$ and all three
fractions were renormalized to sum to 1.0. Feature matrices were standardized
prior to PCA. PVCA was computed both globally (all slides) and per cancer type
(TSS variance within each cancer type).

Silhouette analysis treated TSS labels as cluster assignments and computed the
mean silhouette score on standardized features (Euclidean distance), with
subsampling to \num{5000} slides for computational efficiency. A silhouette
score near zero or negative indicated minimal TSS-driven clustering. Scores
above 0.25 were flagged as moderate batch effects, and scores above 0.5 as
strong batch effects. Per-batch mean silhouette scores identified specific sites
with anomalous feature distributions. We visually inspected UMAP embeddings
colored by TSS to confirm that slides did not cluster by source site after
controlling for cancer type.

\subsection{Power analysis and evidence badges}
\label{sec:methods:power}

We computed the minimum detectable effect size (MDES) at 80\% power
($\alpha = 0.05$, two-sided) for every analysis to characterize the sensitivity
limits of each cancer-type cohort. For survival associations, MDES used the
Schoenfeld--Freedman approximation~\cite{schoenfeld1983, freedman1982}; for
correlations, the Fisher $z$-transform; for categorical associations,
simulation-based power curves (\num{2000} simulated datasets with binary-search
convergence). Formulas are provided in Supplementary Methods.

Every association in the atlas was assigned an evidence-strength badge:
\textbf{strong} ($P_{\text{adj}} < 0.01$, effect size above threshold, narrow
CI, $n \geq 100$), \textbf{moderate} ($P_{\text{adj}} < 0.05$, effect size
above threshold, narrow or moderate CI, $n \geq 50$), \textbf{suggestive}
($P_{\text{adj}} < 0.10$ or CI excludes null [HR $= 1$, $\rho = 0$,
$\delta = 0$, or $\eta^2 = 0$], $n \geq 30$), or \textbf{insufficient}
($n < 30$ or missing statistics). Effect-size thresholds were set at two tiers
guided by conventional benchmarks~\cite{chen2010hazard}.
\textbf{Strong} thresholds (Cohen's medium): HR $\geq 1.5$ (or $\leq 0.667$),
$|\rho| \geq 0.3$, $|\text{Cliff's }\delta| \geq 0.3$, $\eta^2 \geq 0.06$.
\textbf{Moderate} thresholds (Cohen's small): HR $\geq 1.18$ (or $\leq 0.847$),
$|\rho| \geq 0.1$, $|\text{Cliff's }\delta| \geq 0.15$, $\eta^2 \geq 0.01$. CI
width was categorized as ``narrow'' (ratio CI $< 2\times$; additive CI width
$< 0.3$), ``moderate'' ($< 4\times$; $< 0.6$), or ``wide'' (otherwise). Sample
size thresholds were $n \geq 100$ for strong, $n \geq 50$ for moderate, and
$n < 30$ for insufficient evidence.

\subsection{Web application}
\label{sec:methods:webapp}

The HistoAtlas web atlas is built using Astro (static site generator) with
React interactive components. All precomputed statistical results, feature
profiles, cluster metadata, and visualization data are serialized as static
JSON files during the build process. No backend computation server is required
at runtime, enabling deployment on static hosting infrastructure. The interface
provides pan-cancer and per-cancer views of UMAP embeddings, feature
distributions, survival associations, molecular correlations, and cluster
profiles. Users can filter by cancer type, feature category, evidence badge, and
statistical significance.

\paragraph{Spatial interpretability.} The interface provides three complementary
visualization layers. First, a tissue compartment map displays a spatial
segmentation overlay per slide showing five tissue compartments mapped to nine
spatial zones (tumor front, tumor core, peritumoral stroma at three distance
bands [0--50~\textmu m, 50--200~\textmu m, >200~\textmu m], necrosis ring,
necrosis, normal epithelium, and background), enabling users to identify any
pixel's anatomical region by hovering (Fig.\,\ref{fig:fig6}a). Second, for each
of the 38 histomic features, the interface displays the top-5 ranked
224~$\times$~224 pixel tiles from the slide, scored by the feature's computation
strategy. Each tile includes exact pixel coordinates on the whole-slide image, a
10~\textmu m scale bar, and a toggleable cell-type prediction overlay showing 14
cell types in distinct colors (Fig.\,\ref{fig:fig6}c,d). Third, bidirectional
navigation links statistical results to the slides and tissue regions that
generated them: from survival hazard ratios and molecular correlations to
feature pages, and from feature pages to specific tiles on specific slides.

\subsection{Implementation and reproducibility}
\label{sec:methods:implementation}

All analyses were implemented in Python~3.11 and orchestrated by a Snakemake
workflow~\cite{moeldner2021snakemake} that defines a directed acyclic graph of
computational dependencies.

\begin{sloppypar}
Key library versions: \texttt{lifelines}~0.29.0,
\texttt{scipy}~1.12.0,
\texttt{scikit-learn}~1.4.0,
\texttt{umap-learn}~0.5.5,
\texttt{gseapy}~1.1.12,
\texttt{statsmodels}~0.14.1,
\texttt{numpy}~1.26.4,
\texttt{pandas}~2.2.0. All random processes (permutation
tests, bootstrap resampling (effect-size CIs), subsampling (cluster stability), $K$-means initialization) used explicit random
seeds.
\end{sloppypar}

\paragraph{Data and code availability.}
All analysis code, precomputed results, and the web
application are available at \url{https://github.com/histoatlas/histoatlas}.
The interactive atlas is accessible at \url{https://histoatlas.com}
(RRID:\href{https://scicrunch.org/resolver/RRID:SCR_028056}{SCR\_028056}). Derived
feature matrices, precomputed statistical results, and model weights will be
deposited in a public repository with a persistent DOI prior to publication. The
GDC file UUIDs for all \num{6745} slides used in this study are listed in the
code repository.

\bibliographystyle{naturemag}
\bibliography{references}


\onecolumn
\setcounter{table}{0}
\renewcommand{\tablename}{Supplementary Table}

\begin{table}[htbp]
\caption{%
  \textbf{Definition of the 40 histomic features extracted per slide (38 used in downstream analyses).}
  Features are organized into five categories:
  (A)~tissue composition, (B)~cell densities, (C)~nuclear morphology and
  kinetics, (D)~spatial organization, and (E)~spatial heterogeneity.
  $\Omega_{\mathrm{T}}$: tumor compartment;
  $\Omega_{\mathrm{S}}$: stromal compartment;
  $B_T^{0\text{-}50}$/$B_S^{0\text{-}50}$: tumor/stromal band within 50\,\textmu m of the tumor-stroma boundary;
  $B_S^{50\text{-}200}$: stromal band 50--200\,\textmu m from boundary;
  $d_T$: signed distance to tumor boundary (\textmu m);
  $L(\partial)$: boundary contact length (mm);
  $\rho_k(R) = n_k(R) / A(R)$: density of cell type $k$ in region $R$.
  Features~3 and~24 carry zero signal (zero variance across all slides) and are excluded from all downstream analyses, reducing the working feature set to 38.
}
\label{tab:feature_definitions}
\vspace{0.5em}
\centering
{\scriptsize
\begin{tabular}{@{}r l l l@{}}
\toprule
No. & Feature & Formula / definition & Unit \\
\midrule
\multicolumn{4}{@{}l}{\textit{(A) Tissue composition}} \\[2pt]
1 & Tumor area fraction & $A(\Omega_{\mathrm{T}}) / A(\Omega)$ & fraction \\
2 & Stroma area fraction & $A(\Omega_{\mathrm{S}}) / A(\Omega)$ & fraction \\
3 & Normal epithelium area fraction$^\dagger$ & $A(\Omega_{\mathrm{Norm}}) / A(\Omega)$ & fraction \\
4 & Eosinophil-neutrophil ratio (peritumoral) & $[n_{\mathrm{Eos}}(B_S^{0\text{-}50}) + \varepsilon] / [n_{\mathrm{Neu}}(B_S^{0\text{-}50}) + \varepsilon]$ & ratio \\
\midrule
\multicolumn{4}{@{}l}{\textit{(B) Cell densities}} \\[2pt]
5 & Intratumoral cancer cell density & $\rho_{\mathrm{TC}}(\Omega_{\mathrm{T}})$ & cells\,mm$^{-2}$ \\
6 & Intratumoral lymphocyte density & $\rho_{\mathrm{Ly}}(\Omega_{\mathrm{T}})$ & cells\,mm$^{-2}$ \\
7 & Stromal lymphocyte density & $\rho_{\mathrm{Ly}}(\Omega_{\mathrm{S}})$ & cells\,mm$^{-2}$ \\
8 & Intratumoral neutrophil density & $\rho_{\mathrm{Neu}}(\Omega_{\mathrm{T}})$ & cells\,mm$^{-2}$ \\
9 & Intratumoral eosinophil density & $\rho_{\mathrm{Eos}}(\Omega_{\mathrm{T}})$ & cells\,mm$^{-2}$ \\
10 & Stromal fibroblast density & $\rho_{\mathrm{Fib}}(\Omega_{\mathrm{S}})$ & cells\,mm$^{-2}$ \\
\midrule
\multicolumn{4}{@{}l}{\textit{(C) Nuclear morphology and kinetics}} \\[2pt]
11 & Tumor nuclear area (median) & $\mathrm{median}(A_i)$ over tumor nuclei & \textmu m$^2$ \\
12 & Tumor pleomorphism index & $\mathrm{IQR}(A_i) / [\mathrm{median}(A_i) + \varepsilon]$ & unitless \\
13 & Tumor nuclear eccentricity (median) & $\mathrm{median}(e_i)$ over tumor nuclei & unitless \\
14 & Tumor nuclear irregularity (median) & $\mathrm{median}(P_i^2 / 4\pi A_i)$ & unitless \\
15 & Tumor nuclear irregularity (IQR) & $\mathrm{IQR}(P_i^2 / 4\pi A_i)$ & unitless \\
16 & Mitotic index (tumor) & $n_{\mathrm{Mit}}(\Omega_{\mathrm{T}}) / n_{\mathrm{TC}}(\Omega_{\mathrm{T}}) \times 10^3$ & per 1\,k TC \\
17 & Apoptotic index (tumor) & $n_{\mathrm{Apop}}(\Omega_{\mathrm{T}}) / n_{\mathrm{TC}}(\Omega_{\mathrm{T}}) \times 10^3$ & per 1\,k TC \\
18 & Apoptosis-mitosis ratio & $[n_{\mathrm{Apop}} + \varepsilon] / [n_{\mathrm{Mit}} + \varepsilon]$ in $\Omega_{\mathrm{T}}$ & ratio \\
\midrule
\multicolumn{4}{@{}l}{\textit{(D) Spatial organization}} \\[2pt]
19 & Largest tumor component share & $\max_j A(C_j) / A(\Omega_{\mathrm{T}})$ & fraction \\
20 & Tumor region solidity & $A(C_{\max}) / A(\mathrm{Hull}(C_{\max}))$ & unitless \\
21 & Tumor-stroma interface density & $L(\partial(\mathrm{T,S})) / A(\Omega_{\mathrm{T}})$ & mm$^{-1}$ \\
22 & Tumor front fraction & $A(B_T^{0\text{-}50}) / A(\Omega_{\mathrm{T}})$ & fraction \\
23 & Tumor-stroma contact fraction & $L(\partial(\mathrm{T,S})) / \textstyle\sum_{c \neq T} L(\partial(\mathrm{T},c))$ & fraction \\
24 & Tumor-normal contact fraction$^\dagger$ & $L(\partial(\mathrm{T,N})) / \textstyle\sum_{c \neq T} L(\partial(\mathrm{T},c))$ & fraction \\
25 & Lymphocyte infiltration ratio (front) & $\rho_{\mathrm{Ly}}(B_T^{0\text{-}50}) / [\rho_{\mathrm{Ly}}(B_S^{0\text{-}50}) + \varepsilon]$ & ratio \\
26 & Myeloid infiltration ratio (front) & $\rho_{\mathrm{Mye}}(B_T^{0\text{-}50}) / [\rho_{\mathrm{Mye}}(B_S^{0\text{-}50}) + \varepsilon]$ & ratio \\
27 & Deep intratumoral lymphocyte fraction & $n_{\mathrm{Ly}}(d_T > 50) / n_{\mathrm{Ly}}(\Omega_{\mathrm{T}})$ & fraction \\
28 & Peritumoral immune richness & No.\ immune types with ${\geq}\,5$ cells in $B_S^{0\text{-}50}$ & count (0--4) \\
29 & Immune desert fraction & $A(\{x \in \Omega_{\mathrm{T}} : d_{\mathrm{Ly}} > 200\,\mu\mathrm{m}\}) / A(\Omega_{\mathrm{T}})$ & fraction \\
30 & Intratumoral myeloid-lymphoid tilt & $[n_{\mathrm{Neu}} + n_{\mathrm{Eos}}] / [n_{\mathrm{Ly}} + n_{\mathrm{Pla}} + \varepsilon]$ in $\Omega_{\mathrm{T}}$ & ratio \\
31 & Interface-normalized immune pressure & $n_{\mathrm{Ly}}(B_S^{0\text{-}50} \cup B_T^{0\text{-}50}) / L(\partial(\mathrm{T,S}))$ & cells\,mm$^{-1}$ \\
32 & Invasion depth (75th pctl) & p75 of $-d_T(p_i)$ for TC in stroma & \textmu m \\
33 & Tumor-fibroblast coupling (front) & Median NN dist., TC to Fib in $B_T^{0\text{-}50}$ & \textmu m \\
34 & Tumor-lymphocyte NN distance (front) & Median NN dist., TC to Ly in $B_T^{0\text{-}50}$ & \textmu m \\
35 & Peritumoral fibroblast enrichment & $\rho_{\mathrm{Fib}}(B_S^{0\text{-}50}) / [\rho_{\mathrm{Fib}}(B_S^{50\text{-}200}) + \varepsilon]$ & ratio \\
36 & Stromal inflammatory tilt & $[n_{\mathrm{Neu}} + n_{\mathrm{Eos}}] / [n_{\mathrm{Ly}} + n_{\mathrm{Pla}} + \varepsilon]$ in $\Omega_{\mathrm{S}}$ & ratio \\
37 & Fibroblast-lymphocyte proximity (stroma) & Median NN dist., Ly to Fib in $\Omega_{\mathrm{S}}$ & \textmu m \\
\midrule
\multicolumn{4}{@{}l}{\textit{(E) Spatial heterogeneity}} \\[2pt]
38 & Tumor cell density heterogeneity & CV of $\rho_{\mathrm{TC}}$ across tumor tiles & CV \\
39 & Lymphocyte density heterogeneity (tumor) & CV of $\rho_{\mathrm{Ly}}$ across tumor tiles & CV \\
40 & Stromal cellularity heterogeneity & CV of total cell density across stromal tiles & CV \\
\bottomrule
\end{tabular}
}

\vspace{0.5em}
{\scriptsize $^\dagger$\,Zero signal across all slides; excluded from downstream analyses (see Methods).
Abbreviations: TC, tumor cells; Ly, lymphocytes; Neu, neutrophils; Eos, eosinophils;
Fib, fibroblasts; Pla, plasmocytes; Mit, mitotic figures; Apop, apoptotic bodies;
Mye, myeloid cells (Neu\,$+$\,Eos); NN, nearest-neighbor; CV, coefficient of variation;
Hull, convex hull; $\varepsilon = 10^{-6}$.}
\end{table}

\clearpage


{\scriptsize
\begin{longtable}{@{}l l p{5.2cm} l p{2.8cm} l p{2.8cm}@{}}
\caption{%
  \textbf{Biological plausibility audit of 60 atlas-derived claims.}
  Each atomic claim was assessed against the published literature and assigned
  an evidence level: WE = well-established ($>$3 independent confirmations);
  SUP = supported (1--3 prior studies consistent); NP = novel, biologically
  plausible (no prior report, mechanistically coherent); NU = novel, uncertain
  (no prior report, mechanism unclear); C = contradicted (apparent
  contradiction, resolved by category distinction).
  Overall: 27 WE (45\%), 15 SUP (25\%), 12 NP (20\%), 5 NU (8\%), 1 C (2\%).
}
\label{tab:plausibility} \\
\toprule
ID & Axis & Claim & Evid. & Key statistic & Cancer & References \\
\midrule
\endfirsthead
\multicolumn{7}{l}{\small\itshape Supplementary Table~2 continued} \\[4pt]
\toprule
ID & Axis & Claim & Evid. & Key statistic & Cancer & References \\
\midrule
\endhead
\midrule
\multicolumn{7}{r}{\small\itshape Continued on next page} \\
\endfoot
\bottomrule
\endlastfoot

1.1 & Immune & Higher intratumoral TIL density $\to$ better OS in BRCA & WE & HR = 0.72 (0.60--0.88), adj. & BRCA & \cite{fridman2012immune, loi2019, denkert2018} \\
1.2 & Immune & Higher intratumoral TIL density $\to$ better OS pan-cancer & WE & HR = 0.87 (0.81--0.93), adj. & PAN & \cite{fridman2012immune, galon2006} \\
1.3 & Immune & Stromal lymphocyte density shows weaker protective effect than intratumoral pan-cancer & SUP & HR = 0.89 (0.83--0.97), adj. & PAN & \cite{denkert2018} \\
1.4 & Immune & Compartment-specific prognostic strength: intratumoral $>$ stromal & NP & IT HR = 0.87 vs S HR = 0.89, adj. & PAN & \cite{fridman2012immune} \\
1.5 & Immune & Immune desert fraction predicts worse OS in LIHC & SUP & HR = 1.29 (1.12--1.48) & LIHC & \cite{galon2019immunoscore} \\
1.6 & Immune & Deep intratumoral lymphocyte fraction protective in HNSC & SUP & HR = 0.79 (0.68--0.91) & HNSC & \cite{galon2006} \\
1.7 & Immune & Interface-normalized immune pressure protective in HNSC (adjusted) & NP & HNSC HR = 0.74 & HNSC & \cite{galon2006} \\
1.8 & Immune & Peritumoral immune richness has no prognostic value & NU & NS all cancers & PAN & \cite{galon2006} \\
1.9 & Immune & Top gene correlates are T-cell markers and checkpoints & WE & $\rho$ = 0.58--0.63 & BRCA & \cite{thorsson2018immune} \\
1.10 & Immune & Stromal TILs correlate with effector/IFN-$\gamma$ genes & SUP & $\rho$ = 0.49--0.53 & BRCA & \cite{helmink2020} \\
1.11 & Immune & B-cell markers among top TIL correlates & SUP & $\rho$(CD79A) = 0.60 & BRCA & \cite{helmink2020, petitprez2020} \\
1.12 & Immune & Immune subtypes associate with morphometric TIL density & WE & $\eta^2$ = 0.12--0.13 & BRCA & \cite{thorsson2018immune} \\

2.1 & Prolif. & High mitotic index $\to$ worse OS pan-cancer & WE & HR = 1.25 (1.19--1.31) & PAN & \cite{elston1991grading} \\
2.2 & Prolif. & CCNE1 is top correlate of mitotic index in BRCA & WE & $\rho$ = 0.57 & BRCA & \\
2.3 & Prolif. & PLK1, AURKA, BIRC5 are canonical mitotic correlates & WE & $\rho$ = 0.52--0.65 & Multi & \cite{nigg2001} \\
2.4 & Prolif. & MKI67 and TOP2A validate morphometric mitotic index & WE & $\rho$ = 0.48--0.60 & Multi & \\
2.5 & Prolif. & Apoptotic index protective in LIHC & SUP & HR = 0.66 (0.55--0.79) & LIHC & \\
2.6 & Prolif. & Apoptosis/mitosis ratio protective pan-cancer & SUP & HR = 0.79 (0.75--0.84) & PAN & \\
2.7 & Prolif. & ACC has extreme cell turnover sensitivity & NP & Ratio HR = 0.36 (0.20--0.66) & ACC & \\
2.8 & Prolif. & APC negatively correlates with mitotic index in STAD & WE & $\rho$ = $-$0.24 & STAD & \\
2.9 & Prolif. & Tumor cell density heterogeneity null pan-cancer & NU & HR $\approx$ 1.00, NS & PAN & \cite{marusyk2012} \\
2.10 & Prolif. & Reversed mitotic index effects in COAD, ESCA, OV & NU & HR $\approx$ 0.78, NS & Multi & \\

3.1 & Nuclear & Larger nuclei $\to$ worse OS pan-cancer & WE & HR = 1.19, $P$ = $5{\times}10^{-12}$ & PAN & \cite{elston1991grading, abel2024} \\
3.2 & Nuclear & Nuclear area predicts OS in HCC & WE & HR = 1.20 & LIHC & \\
3.3 & Nuclear & Pleomorphism correlates with PLK1, AURKA, CCNE1, MKI67 & WE & $\rho$ = 0.40--0.49 & BRCA & \cite{nigg2001} \\
3.4 & Nuclear & Pleomorphism inversely correlates with BCL2 and ESR1 & WE & $\rho$ = $-$0.36 to $-$0.37 & BRCA & \\
3.5 & Nuclear & Nuclear eccentricity has opposing tissue-specific effects & NP & UCEC HR = 0.70; LIHC HR = 1.32 & Multi & \cite{abel2024} \\
3.6 & Nuclear & Nuclear irregularity IQR predicts OS only in LIHC & NP & HR = 1.41 & LIHC & \\
3.7 & Nuclear & Nuclear irregularity protective in HNSC & NU & HR = 0.78 & HNSC & \\

4.1 & Invasion & Greater invasion depth $\to$ worse OS pan-cancer & WE & HR = 1.11 & PAN & \\
4.2 & Invasion & Invasion depth correlates with EMT markers & WE & $\rho$(ZEB1) = 0.32 & BRCA & \cite{nieto2016emt} \\
4.3 & Invasion & Invasion depth correlates with ALDH1A1 & SUP & $\rho$ = 0.21 & BRCA & \\
4.4 & Invasion & TGFB1 correlates with invasion in BRCA and PAAD & WE & $\rho$ = 0.29 & Multi & \cite{nieto2016emt} \\
4.5 & Invasion & Invasion depth inversely correlates with proliferation & SUP & $\rho$(E2F targets) = $-$0.30 & BRCA & \cite{hatzikirou2012, giese2003glioma} \\
4.6 & Invasion & Tumor--stroma interface density protective pan-cancer & NP & HR = 0.85, $P$ = $1{\times}10^{-8}$ & PAN & \\
4.7 & Invasion & Fibroblast coupling at front predicts OS in LIHC & SUP & HR = 1.48, $P$ = $7{\times}10^{-5}$ & LIHC & \cite{sahai2020} \\

5.1 & Stromal & Eosinophil infiltration protective in BRCA and HNSC & WE & BRCA HR = 0.63; HNSC HR = 0.77 & Multi & \cite{carretero2015} \\
5.2 & Stromal & Eosinophil density correlates with cytotoxic T-cell signatures & WE & $\rho$(GZMB) = 0.40 & BRCA & \cite{carretero2015} \\
5.3 & Stromal & Neutrophil density is context-dependent & WE & NS pan-cancer & PAN & \\
5.4 & Stromal & Eosinophil/neutrophil ratio captures innate immune polarization & NP & BRCA HR = 0.65 & BRCA & \\
5.5 & Stromal & Fibroblast density non-prognostic pan-cancer & SUP & HR = 0.97 & PAN & \cite{sahai2020} \\
5.6 & Stromal & Stromal inflammatory tilt non-prognostic & NU & NS all cancers & PAN & \\

6.1 & Spatial & Lymphocyte density at invasive front predicts improved OS & WE & HR = 0.85, $P$ = $1{\times}10^{-8}$ & PAN & \cite{galon2006, pages2018} \\
6.2 & Spatial & NN distance captures spatial immune exclusion & SUP & $\rho$(cytotoxic) = $-$0.54 & BRCA & \cite{schurch2020} \\
6.3 & Spatial & Greater NN distance $\to$ worse gene expression signatures & WE & $\rho$(CD8A) = $-$0.53 & BRCA & \cite{chen2017oncology, schurch2020} \\
6.4 & Spatial & Myeloid-to-lymphoid tilt adversely prognostic & SUP & HR = 1.10, $P$ = $3{\times}10^{-5}$ & PAN & \cite{thorsson2018immune} \\
6.5 & Spatial & Lymphocyte infiltration at front correlates with B-cell signatures & SUP & $\rho$(B-cell) = 0.49 & BRCA & \cite{helmink2020, petitprez2020} \\
6.6 & Spatial & Myeloid infiltration at front independently protective & NP & HR = 0.92 & PAN & \\
6.7 & Spatial & NN distance at front associated with worse OS in LUAD (unadjusted only) & NP & HR = 1.29 (unadj); NS adjusted & LUAD & \\

7.1 & Tissue & Tumor area fraction is a proxy for tumor purity & WE & $\rho$(prolif) = 0.40 & PAN & \cite{aran2015} \\
7.2 & Tissue & Normal epithelium near-zero in resected tumors & WE & -- & PAN & \\
7.3 & Tissue & Lymphocyte density heterogeneity (spatial CV) is protective & NP & HR = 0.74 (LIHC); HR = 0.82 (HNSC) & Multi & \cite{sautesfridman2019tls} \\
7.4 & Tissue & Stromal cellularity heterogeneity protective in UCEC & NP & HR = 0.61 & UCEC & \\
7.5 & Tissue & Morphological heterogeneity is protective (contradicts ITH paradigm) & C & Protective in 11/15 cancers (adjusted) & PAN & \cite{marusyk2012} \\
7.6 & Tissue & Tumor area fraction non-prognostic pan-cancer & WE & HR = 1.01, NS & PAN & \cite{aran2015} \\

8.1 & Cluster & Immune-hot clusters enriched for C2 (IFN-$\gamma$) & WE & Clusters 3, 4, 7 all C2+ & PAN & \cite{thorsson2018immune} \\
8.2 & Cluster & Proliferative clusters have worse survival & WE & $r \approx 0.85$ across clusters & PAN & \cite{elston1991grading} \\
8.3 & Cluster & Tissue-specific clusters recapitulate organ pathways & WE & THYM $\to$ immune rejection; CRC $\to$ Wnt/$\beta$-catenin & PAN & \cite{radovich2018thymoma, tcga2012crc} \\
8.4 & Cluster & Mutation enrichment reflects TMB, not specific drivers & SUP & Binary enrich/deplete pattern & PAN & \\
8.5 & Cluster & Two distinct immune-cold phenotypes with opposite survival & NP & C2: HR = 0.54; C8: HR = 1.37 & PAN & \\

\end{longtable}
}

\clearpage


\begin{table}[htbp]
\caption{%
  \textbf{TCGA cancer types included and excluded from the HistoAtlas analysis.}
  Twenty-one solid-tumor cancer types were included. Twelve additional cancer
  types were excluded because their dominant cell morphologies fall outside
  the training domain of the cell detection model.
}
\label{tab:cancer_types}
\vspace{0.5em}
\centering
{\scriptsize
\begin{tabular}{@{}l l r l@{}}
\toprule
Abbreviation & Full name & $N$ slides & Status \\
\midrule
ACC & Adrenocortical carcinoma & 227 & Included \\
BLCA & Bladder urothelial carcinoma & 417 & Included \\
BRCA & Breast invasive carcinoma & \num{1037} & Included \\
CESC & Cervical squamous cell carcinoma & 279 & Included \\
CHOL & Cholangiocarcinoma & 38 & Included \\
COAD & Colon adenocarcinoma & 441 & Included \\
ESCA & Esophageal carcinoma & 158 & Included \\
HNSC & Head and neck squamous cell carcinoma & 471 & Included \\
LIHC & Liver hepatocellular carcinoma & 365 & Included \\
LUAD & Lung adenocarcinoma & 511 & Included \\
LUSC & Lung squamous cell carcinoma & 357 & Included \\
MESO & Mesothelioma & 82 & Included \\
OV & Ovarian serous cystadenocarcinoma & 107 & Included \\
PAAD & Pancreatic adenocarcinoma & 146 & Included \\
PRAD & Prostate adenocarcinoma & 353 & Included \\
READ & Rectum adenocarcinoma & 157 & Included \\
STAD & Stomach adenocarcinoma & 400 & Included \\
THCA & Thyroid carcinoma & 473 & Included \\
THYM & Thymoma & 180 & Included \\
UCEC & Uterine corpus endometrial carcinoma & 459 & Included \\
UCS & Uterine carcinosarcoma & 87 & Included \\
\midrule
DLBC & Diffuse large B-cell lymphoma & -- & Excluded (lymphoid cells) \\
GBM & Glioblastoma multiforme & -- & Excluded (glial cells) \\
KICH & Kidney chromophobe & -- & Excluded (renal tubular) \\
KIRC & Kidney renal clear cell carcinoma & -- & Excluded (renal tubular) \\
KIRP & Kidney renal papillary cell carcinoma & -- & Excluded (renal tubular) \\
LAML & Acute myeloid leukemia & -- & Excluded (myeloid blasts) \\
LGG & Brain lower grade glioma & -- & Excluded (glial cells) \\
PCPG & Pheochromocytoma and paraganglioma & -- & Excluded (neuroendocrine) \\
SARC & Sarcoma & -- & Excluded (mesenchymal) \\
SKCM & Skin cutaneous melanoma & -- & Excluded (melanocytes) \\
TGCT & Testicular germ cell tumors & -- & Excluded (germ cells) \\
UVM & Uveal melanoma & -- & Excluded (melanocytes) \\
\bottomrule
\end{tabular}
}
\end{table}

\clearpage


\begin{table}[htbp]
\caption{%
  \textbf{Curated gene panel (133 genes) used for molecular correlation analysis.}
  Genes were selected from established cancer gene panels, immune checkpoint
  targets, and EMT/stemness markers. Functional categories are provided for
  annotation; genes may participate in multiple pathways.
}
\label{tab:gene_panel}
\vspace{0.5em}
\centering
{\scriptsize
\begin{tabular}{@{}l p{12cm}@{}}
\toprule
Category & Genes \\
\midrule
Immune / checkpoint (36) & CD274 (PD-L1), PDCD1 (PD-1), PDCD1LG2 (PD-L2), CTLA4, LAG3, TIGIT, HAVCR2 (TIM-3), IDO1, CD8A, CD8B, CD4, CD3D, CD3E, FOXP3, CD19, CD79A, MS4A1 (CD20), CD14, CD68, CD163, CD40, CD80, CD86, ITGAM, NKG7, IFNG, GZMA, GZMB, PRF1, TNF, IL6, IL10, IL2, CXCL9, CXCL10, TGFB1 \\
Proliferation / cell cycle (15) & MKI67, TOP2A, PCNA, CCNB1, CCND1, CCNE1, CDK1, CDK2, CDK4, CDK6, PLK1, AURKA, BIRC5, MCM2, E2F1 \\
EMT / invasion / stemness (15) & CDH1, CDH2, VIM, SNAI1, SNAI2, ZEB1, ZEB2, TWIST1, FN1, ACTA2, ALDH1A1, CD44, SOX2, NANOG, PROM1 \\
Apoptosis / DNA damage (19) & TP53, BCL2, BCL2L1, MCL1, BAX, CASP3, CASP8, FAS, BRCA1, BRCA2, ATM, CHEK1, CHEK2, RAD51, PARP1, MLH1, MSH2, MSH6, CDKN2A \\
Signaling / oncogenes (27) & EGFR, ERBB2 (HER2), MET, KRAS, NRAS, HRAS, BRAF, NF1, PIK3CA, PTEN, AKT1, MTOR, NOTCH1, FBXW7, CTNNB1, APC, SMAD4, RB1, MYC, ALK, RET, FGFR1, FGFR2, FGFR3, KIT, PDGFRA, JAK2 \\
Hormone receptors (3) & ESR1, PGR, AR \\
Epigenetic / chromatin (6) & ARID1A, SMARCA4, IDH1, IDH2, SETD2, BAP1 \\
Metabolism / hypoxia (5) & VEGFA, HIF1A, LDHA, SLC2A1, CA9 \\
Other (7) & NFE2L2, KEAP1, STK11, NF2, VHL, MGMT, TERT \\
\bottomrule
\end{tabular}
}
\vspace{0.3em}

{\scriptsize Gene names follow HUGO Gene Nomenclature Committee (HGNC) conventions. Common aliases are shown in parentheses.}
\end{table}

\clearpage


\begin{table}[htbp]
\caption{%
  \textbf{50 MSigDB Hallmark pathways used for ssGSEA scoring.}
  Pathway activity scores were computed via single-sample Gene Set Enrichment
  Analysis (ssGSEA)~\cite{barbie2009ssgsea, hanzelmann2013gsva} using the 50 Hallmark gene sets
  from MSigDB~\cite{liberzon2015} (collection identifier: \texttt{h.all},
  available at \url{https://www.gsea-msigdb.org/gsea/msigdb/collection_details.jsp}).
  Each Hallmark gene set contains ${\sim}$200 genes (range: 32--200) and represents
  a well-characterized biological process or state. Gene sets with fewer than
  10 genes present in the expression matrix after intersection were excluded.
  Full gene lists for all 50 Hallmark pathways are available from MSigDB
  (Liberzon et~al., 2015).
}
\label{tab:pathway_signatures}
\vspace{0.5em}
\centering
{\scriptsize
\begin{tabular}{@{}l l@{}}
\toprule
MSigDB Hallmark pathway & Biological category \\
\midrule
HALLMARK\_ADIPOGENESIS & Metabolic \\
HALLMARK\_ALLOGRAFT\_REJECTION & Immune \\
HALLMARK\_ANDROGEN\_RESPONSE & Hormonal \\
HALLMARK\_ANGIOGENESIS & Development \\
HALLMARK\_APICAL\_JUNCTION & Cellular component \\
HALLMARK\_APICAL\_SURFACE & Cellular component \\
HALLMARK\_APOPTOSIS & Proliferation \\
HALLMARK\_BILE\_ACID\_METABOLISM & Metabolic \\
HALLMARK\_CHOLESTEROL\_HOMEOSTASIS & Metabolic \\
HALLMARK\_COAGULATION & Immune \\
HALLMARK\_COMPLEMENT & Immune \\
HALLMARK\_DNA\_REPAIR & DNA damage \\
HALLMARK\_E2F\_TARGETS & Proliferation \\
HALLMARK\_EPITHELIAL\_MESENCHYMAL\_TRANSITION & Development \\
HALLMARK\_ESTROGEN\_RESPONSE\_EARLY & Hormonal \\
HALLMARK\_ESTROGEN\_RESPONSE\_LATE & Hormonal \\
HALLMARK\_FATTY\_ACID\_METABOLISM & Metabolic \\
HALLMARK\_G2M\_CHECKPOINT & Proliferation \\
HALLMARK\_GLYCOLYSIS & Metabolic \\
HALLMARK\_HEDGEHOG\_SIGNALING & Signaling \\
HALLMARK\_HEME\_METABOLISM & Metabolic \\
HALLMARK\_HYPOXIA & Metabolic \\
HALLMARK\_IL2\_STAT5\_SIGNALING & Immune \\
HALLMARK\_IL6\_JAK\_STAT3\_SIGNALING & Immune \\
HALLMARK\_INFLAMMATORY\_RESPONSE & Immune \\
HALLMARK\_INTERFERON\_ALPHA\_RESPONSE & Immune \\
HALLMARK\_INTERFERON\_GAMMA\_RESPONSE & Immune \\
HALLMARK\_KRAS\_SIGNALING\_DN & Signaling \\
HALLMARK\_KRAS\_SIGNALING\_UP & Signaling \\
HALLMARK\_MITOTIC\_SPINDLE & Proliferation \\
HALLMARK\_MTORC1\_SIGNALING & Signaling \\
HALLMARK\_MYC\_TARGETS\_V1 & Proliferation \\
HALLMARK\_MYC\_TARGETS\_V2 & Proliferation \\
HALLMARK\_MYOGENESIS & Development \\
HALLMARK\_NOTCH\_SIGNALING & Signaling \\
HALLMARK\_OXIDATIVE\_PHOSPHORYLATION & Metabolic \\
HALLMARK\_P53\_PATHWAY & Proliferation \\
HALLMARK\_PANCREAS\_BETA\_CELLS & Development \\
HALLMARK\_PEROXISOME & Metabolic \\
HALLMARK\_PI3K\_AKT\_MTOR\_SIGNALING & Signaling \\
HALLMARK\_PROTEIN\_SECRETION & Cellular component \\
HALLMARK\_REACTIVE\_OXYGEN\_SPECIES\_PATHWAY & Metabolic \\
HALLMARK\_SPERMATOGENESIS & Development \\
HALLMARK\_TGF\_BETA\_SIGNALING & Signaling \\
HALLMARK\_TNFA\_SIGNALING\_VIA\_NFKB & Immune \\
HALLMARK\_UNFOLDED\_PROTEIN\_RESPONSE & Cellular component \\
HALLMARK\_UV\_RESPONSE\_DN & DNA damage \\
HALLMARK\_UV\_RESPONSE\_UP & DNA damage \\
HALLMARK\_WNT\_BETA\_CATENIN\_SIGNALING & Signaling \\
HALLMARK\_XENOBIOTIC\_METABOLISM & Metabolic \\
\bottomrule
\end{tabular}
}
\end{table}

\clearpage


\begin{table}[htbp]
\caption{%
  \textbf{Benjamini--Hochberg correction family definitions by analysis type.}
  Each family groups biologically coherent tests to control the false discovery
  rate within meaningful contexts. Every adjusted $P$-value is stored alongside
  its correction family identifier and the number of tests in the family.
}
\label{tab:bh_families}
\vspace{0.5em}
\centering
{\scriptsize
\begin{tabular}{@{}l l l@{}}
\toprule
Analysis type & Family definition & Typical family size \\
\midrule
Survival (Cox) & Cancer type $\times$ endpoint $\times$ model & 36--38 \\
Survival (RMST) & Cancer type $\times$ endpoint $\times$ model & 36--38 \\
Molecular correlations & Cancer type $\times$ target set $\times$ method $\times$ model & 133--293 \\
Categorical associations & Cancer type $\times$ variable $\times$ test type $\times$ model & 36--38 \\
Cluster survival (Cox) & Cluster level $\times$ analysis type $\times$ cancer $\times$ endpoint $\times$ model & 10--69 \\
Cluster survival (log-rank) & Cluster level $\times$ analysis type $\times$ cancer $\times$ endpoint & 10--69 \\
Cluster enrichment (mutation) & Cluster level $\times$ cancer type & 50--133 \\
Cluster enrichment (pathway) & Cluster level $\times$ cancer type & 50 \\
GSEA & Canonical FDR (pooled null NES) & All gene sets \\
\bottomrule
\end{tabular}
}
\end{table}

\clearpage


\begin{table}[htbp]
\caption{%
  \textbf{Proportional hazards assumption violation prevalence.}
  For each cancer type and survival endpoint, the fraction of fitted Cox models
  classified as ``pass'' ($P \geq 0.05$), ``warn'' ($0.01 \leq P < 0.05$), or
  ``fail'' ($P < 0.01$) based on the Schoenfeld residual test. When the PH
  assumption failed, restricted mean survival time (RMST) was computed as a
  complementary summary. Values shown for the adjusted model (age, sex, stage;
  stratified by tissue source site); unadjusted models show similar patterns.
}
\label{tab:ph_violations}
\vspace{0.5em}
\centering
\vspace{0.5em}
\centering
{\scriptsize
\begin{tabular}{@{}l r r r r@{}}
\toprule
Cancer type & $N$ models & Pass (\%) & Warn (\%) & Fail (\%) \\
\midrule
ACC & 114 & 112 (98) & 2 (2) & 0 (0) \\
BLCA & 152 & 145 (95) & 7 (5) & 0 (0) \\
BRCA & 152 & 141 (93) & 9 (6) & 2 (1) \\
CESC & 152 & 135 (89) & 8 (5) & 9 (6) \\
CHOL & 114 & 109 (96) & 5 (4) & 0 (0) \\
COAD & 152 & 145 (95) & 4 (3) & 3 (2) \\
ESCA & 152 & 142 (93) & 9 (6) & 1 (1) \\
HNSC & 152 & 144 (95) & 6 (4) & 2 (1) \\
LIHC & 152 & 140 (92) & 5 (3) & 7 (5) \\
LUAD & 152 & 145 (95) & 6 (4) & 1 (1) \\
LUSC & 152 & 135 (89) & 12 (8) & 5 (3) \\
MESO & 114 & 112 (98) & 2 (2) & 0 (0) \\
OV & 152 & 142 (93) & 7 (5) & 3 (2) \\
PAAD & 152 & 151 (99) & 1 (1) & 0 (0) \\
PRAD & 76 & 69 (91) & 4 (5) & 3 (4) \\
READ & 114 & 113 (99) & 1 (1) & 0 (0) \\
STAD & 152 & 136 (89) & 13 (9) & 3 (2) \\
THCA & 113 & 110 (97) & 3 (3) & 0 (0) \\
THYM & 38 & 35 (92) & 2 (5) & 1 (3) \\
UCEC & 152 & 147 (97) & 5 (3) & 0 (0) \\
\midrule
PANCAN & 152 & 107 (70) & 16 (11) & 29 (19) \\
\midrule
\textbf{Total} & \textbf{\num{2811}} & \textbf{\num{2615} (93.0)} & \textbf{127 (4.5)} & \textbf{69 (2.5)} \\
\bottomrule
\end{tabular}
}
\vspace{0.3em}

{\scriptsize
$N$ models: number of adjusted Cox regressions (features $\times$ endpoints
available for each cancer type). UCS is excluded because the adjusted model
did not converge for this cohort.
Not all endpoints are evaluable in all cancer
types; see Supplementary Table~\ref{tab:rmst_horizons} for horizon definitions.
Pass: Schoenfeld $P \geq 0.05$; Warn: $0.01 \leq P < 0.05$;
Fail: $P < 0.01$. Unadjusted models show similar patterns.
The pan-cancer cohort exhibits the highest violation rate because cancer-type
heterogeneity introduces non-proportional baseline hazards.
}
\end{table}

\clearpage


\begin{table}[htbp]
\caption{%
  \textbf{Sample sizes per cancer type and model tier for overall survival.}
  Sample sizes for the unadjusted (feature only) and adjusted
  (feature + age + sex + stage; TSS-stratified) models. When MICE imputation
  was used, sample sizes reflect the imputed dataset; differences between
  tiers reflect cancer types where the adjusted model was not fitted due to
  insufficient covariate data.
}
\label{tab:complete_case_n}
\vspace{0.5em}
\centering
{\scriptsize
\begin{tabular}{@{}l r r l@{}}
\toprule
Cancer type & $n$ (unadjusted) & $n$ (adjusted) & MICE used \\
\midrule
ACC  &  55 &  55 & Yes \\
BLCA & 348 & 348 & Yes \\
BRCA & 960 & 960 & Yes \\
CESC & 261 & 261 & No$^\ddagger$  \\
CHOL &  36 &  36 & No  \\
COAD & 413 & 413 & Yes \\
ESCA & 155 & 155 & Yes \\
HNSC & 444 & 444 & Yes \\
LIHC & 353 & 353 & Yes \\
LUAD & 441 & 441 & Yes \\
LUSC & 322 & 322 & Yes \\
MESO &  70 &  70 & No  \\
OV   & 103 & 103 & No$^\ddagger$  \\
PAAD & 134 & 134 & Yes \\
READ & 141 & 141 & Yes \\
STAD & 369 & 369 & Yes \\
THCA & 455 & 455 & Yes \\
UCEC & 409 & 409 & Yes$^\ddagger$ \\
UCS  &  53 & -- & -- \\
\midrule
PANCAN & \num{5957} & \num{4560} & No \\
\bottomrule
\end{tabular}
}
\vspace{0.3em}

{\scriptsize
PRAD and THYM are omitted because OS was not evaluable for these cancer
types (see Supplementary Table~\ref{tab:rmst_horizons}).
UCS is marked ``--'' because the adjusted model did not converge for
this cohort.
For cancer types where MICE was used, the $n$~(adjusted) reflects the
imputed dataset (5~imputations, pooled via Rubin's rules); the
$n$~(unadjusted) column uses complete cases for the feature and
survival outcome only.
$^\ddagger$\,CESC, OV, and UCEC use age and sex only as covariates
(pathologic stage was unavailable); all other adjusted models include
age, sex, and stage.
PANCAN uses complete-case analysis (no MICE); the reduced $n$~(adjusted)
reflects exclusion of cases with missing stage data.
}
\end{table}

\clearpage


\begin{table}[htbp]
\caption{%
  \textbf{Cancer-type-specific RMST time horizons.}
  The restricted mean survival time (RMST) was computed using cancer-type-specific
  truncation horizons chosen to ensure adequate follow-up and at-risk populations.
}
\label{tab:rmst_horizons}
\vspace{0.5em}
\centering
{\scriptsize
\begin{tabular}{@{}l r l@{}}
\toprule
Cancer type & Horizon (days) & Rationale \\
\midrule
PAAD & 730 (2 yr) & Aggressive; short median survival \\
MESO & 730 (2 yr) & Aggressive; short median survival \\
THCA & 1825 (5 yr) & Indolent; long median survival \\
PRAD & 1825 (5 yr) & Indolent; long median survival \\
All others & 1095 (3 yr) & Default \\
\bottomrule
\end{tabular}
}
\end{table}


\begin{table}[htbp]
\caption{%
  \textbf{Descriptive statistics for the 38 histomic features across
  \num{6745} slides.}
  Feature numbering follows Supplementary Table~1; features~3 and~24
  (zero variance) are excluded.
  Cell density, nearest-neighbor distance, and heterogeneity features are
  reported on a $\log_e$-transformed scale (see Methods,
  \S\ref{sec:methods:preprocessing}).
  IQR: interquartile range (Q1--Q3).
}
\label{tab:feature_stats}
\vspace{0.5em}
\centering
{\scriptsize
\begin{tabular}{@{}r l r r r r r r@{}}
\toprule
No. & Feature & Min & Max & Mean & Std & Median & IQR \\
\midrule
\multicolumn{8}{@{}l}{\textit{(A) Tissue composition}} \\[2pt]
1  & Tumor area fraction          & 0.04 & 0.93 & 0.45 & 0.20 & 0.43 & 0.29--0.60 \\
2  & Stroma area fraction         & 0.02 & 0.92 & 0.45 & 0.22 & 0.46 & 0.28--0.62 \\
4  & Eos/neu ratio (peritumoral)  & 0.00 & 2.70 & 0.40 & 0.47 & 0.22 & 0.09--0.52 \\
\midrule
\multicolumn{8}{@{}l}{\textit{(B) Cell densities (log scale)}} \\[2pt]
5  & Cancer cell density (IT)     & 5.15 & 9.39 & 8.37 & 0.58 & 8.47 & 8.10--8.75 \\
6  & Lymphocyte density (IT)      & 2.52 & 10.01 & 6.03 & 1.40 & 6.00 & 5.08--6.96 \\
7  & Lymphocyte density (S)       & 3.81 & 7.92 & 6.11 & 0.82 & 6.17 & 5.56--6.73 \\
8  & Neutrophil density (IT)      & 0.00 & 6.87 & 2.39 & 1.58 & 2.14 & 1.09--3.47 \\
9  & Eosinophil density (IT)      & 0.00 & 5.25 & 1.41 & 1.31 & 0.98 & 0.30--2.32 \\
10 & Fibroblast density (S)       & 5.69 & 8.94 & 7.41 & 0.53 & 7.42 & 7.09--7.74 \\
\midrule
\multicolumn{8}{@{}l}{\textit{(C) Nuclear morphology and kinetics}} \\[2pt]
11 & Nuclear area (median)        & 16.59 & 72.24 & 38.72 & 9.93 & 37.50 & 31.75--44.75 \\
12 & Pleomorphism index           & 0.33 & 1.19 & 0.67 & 0.15 & 0.66 & 0.58--0.76 \\
13 & Nuclear eccentricity         & 0.52 & 0.81 & 0.71 & 0.05 & 0.72 & 0.69--0.75 \\
14 & Nuclear irregularity (median)& 1.06 & 1.33 & 1.16 & 0.05 & 1.15 & 1.13--1.19 \\
15 & Nuclear irregularity (IQR)   & 0.05 & 0.45 & 0.18 & 0.07 & 0.17 & 0.14--0.21 \\
16 & Mitotic index                & 0.00 & 2.42 & 0.75 & 0.66 & 0.61 & 0.13--1.27 \\
17 & Apoptotic index              & 0.99 & 6.61 & 3.35 & 0.93 & 3.33 & 2.76--3.89 \\
18 & Apoptosis/mitosis ratio      & 0.82 & 8.12 & 3.59 & 1.61 & 3.31 & 2.33--4.66 \\
\midrule
\multicolumn{8}{@{}l}{\textit{(D) Spatial organization}} \\[2pt]
19 & Largest tumor component      & 0.01 & 0.99 & 0.32 & 0.26 & 0.24 & 0.11--0.46 \\
20 & Tumor region solidity        & 0.21 & 0.84 & 0.50 & 0.13 & 0.49 & 0.41--0.58 \\
21 & T--S interface density       & 0.46 & 89.50 & 25.59 & 18.19 & 21.85 & 11.64--35.66 \\
22 & Tumor front fraction         & 0.04 & 1.00 & 0.54 & 0.26 & 0.55 & 0.32--0.76 \\
23 & T--S contact fraction        & 0.01 & 0.80 & 0.34 & 0.18 & 0.34 & 0.20--0.47 \\
25 & Ly infiltration ratio (front)& 0.09 & 2.16 & 0.73 & 0.40 & 0.67 & 0.44--0.95 \\
26 & Myeloid infilt.\ ratio (front)& 0.03 & 1.76 & 0.44 & 0.27 & 0.39 & 0.25--0.56 \\
27 & Deep IT lymphocyte fraction  & 0.00 & 0.96 & 0.39 & 0.24 & 0.35 & 0.19--0.56 \\
28 & Peritumoral immune richness  & 0.02 & 0.88 & 0.34 & 0.22 & 0.29 & 0.15--0.52 \\
29 & Immune desert fraction       & 0.00 & 0.41 & 0.03 & 0.06 & 0.01 & 0.00--0.03 \\
30 & IT myeloid--lymphoid tilt    & 0.00 & 1.11 & 0.08 & 0.16 & 0.02 & 0.01--0.08 \\
31 & Interface immune pressure    & 0.77 & 6.34 & 3.21 & 1.00 & 3.23 & 2.53--3.86 \\
32 & Invasion depth (p75)         & $-$0.76 & 123.70 & 33.77 & 21.16 & 29.28 & 20.56--41.34 \\
33 & TC--Fib coupling (front)     & 2.79 & 4.81 & 3.56 & 0.36 & 3.53 & 3.31--3.76 \\
34 & TC--Ly NN distance (front)   & 1.96 & 5.71 & 3.95 & 0.56 & 3.96 & 3.65--4.26 \\
35 & Peritumoral Fib enrichment   & 0.38 & 1.28 & 0.72 & 0.14 & 0.71 & 0.63--0.79 \\
36 & Stromal inflammatory tilt    & 0.00 & 1.06 & 0.12 & 0.17 & 0.06 & 0.02--0.16 \\
37 & Fib--Ly proximity (stroma)   & 2.04 & 3.82 & 2.67 & 0.26 & 2.65 & 2.52--2.80 \\
\midrule
\multicolumn{8}{@{}l}{\textit{(E) Spatial heterogeneity (log scale)}} \\[2pt]
38 & TC density heterogeneity     & 0.00 & 9.32 & 7.94 & 1.13 & 8.06 & 7.69--8.46 \\
39 & Ly density heterogeneity (IT)& 0.00 & 9.45 & 4.63 & 2.77 & 5.37 & 4.44--6.34 \\
40 & Stromal cell.\ heterogeneity & 6.46 & 9.01 & 7.68 & 0.45 & 7.66 & 7.38--7.97 \\
\bottomrule
\end{tabular}
}
\vspace{0.3em}

{\scriptsize
Abbreviations: IT, intratumoral; S, stromal; T--S, tumor--stroma;
Ly, lymphocyte; TC, tumor cell; Fib, fibroblast; NN, nearest-neighbor;
Eos, eosinophil; Neu, neutrophil; CV, coefficient of variation.
}
\end{table}

\clearpage


{\scriptsize
\begin{longtable}{@{}l r r@{/}r r@{/}r r@{/}r r@{/}r r@{/}r r@{}}
\caption{%
  \textbf{Significant Spearman correlations per cancer type and molecular
  data type.}
  Each cell shows the number of significant correlations (false discovery rate~$< 0.05$
  after Benjamini--Hochberg correction within predefined families) out of
  the total tested.
  Totals include all adjustment models.
  Significance rates vary with sample size: PANCAN ($n = \num{4654}$) and
  BRCA ($n = 953$) show the highest yield, while small cohorts (UCS,
  $n = 53$; CHOL, $n = 36$) have low power.
}
\label{tab:correlation_breakdown} \\
\toprule
Cancer & $n$ & \multicolumn{2}{c}{Expression} & \multicolumn{2}{c}{CNV} & \multicolumn{2}{c}{Pathway} & \multicolumn{2}{c}{Immune} & \multicolumn{2}{c}{Total} & \% \\
\midrule
\endfirsthead
\multicolumn{13}{l}{\small\itshape Supplementary Table~11 continued} \\[4pt]
\toprule
Cancer & $n$ & \multicolumn{2}{c}{Expression} & \multicolumn{2}{c}{CNV} & \multicolumn{2}{c}{Pathway} & \multicolumn{2}{c}{Immune} & \multicolumn{2}{c}{Total} & \% \\
\midrule
\endhead
\midrule
\multicolumn{13}{r}{\small\itshape Continued on next page} \\
\endfoot
\bottomrule
\endlastfoot
ACC    & 52 & 370 & \num{10640} & 0 & \num{10640} & 139 & \num{4000} & 62 & 480 & 571 & \num{25760} & 2.2 \\
BLCA   & 344 & \num{3916} & \num{10640} & 421 & \num{10640} & \num{1718} & \num{4000} & 299 & 480 & \num{6354} & \num{25760} & 24.7 \\
BRCA   & 953 & \num{6547} & \num{10906} & \num{2967} & \num{10906} & \num{2735} & \num{4100} & 346 & 492 & \num{12595} & \num{26404} & 47.7 \\
CESC   & 259 & \num{1052} & \num{5320} & 1 & \num{5320} & 444 & \num{2000} & 83 & 240 & \num{1580} & \num{12880} & 12.3 \\
CHOL   & 36 & 168 & \num{10640} & 0 & \num{10640} & 0 & 0 & 39 & 480 & 207 & \num{21760} & 1.0 \\
COAD   & 405 & \num{2856} & \num{10640} & 429 & \num{10640} & \num{1214} & \num{4000} & 238 & 480 & \num{4737} & \num{25760} & 18.4 \\
ESCA   & 141 & 890 & \num{10374} & 0 & \num{10374} & 254 & \num{3900} & 92 & 468 & \num{1236} & \num{25116} & 4.9 \\
HNSC   & 412 & \num{2889} & \num{10640} & 843 & \num{10640} & \num{1069} & \num{4000} & 185 & 480 & \num{4986} & \num{25760} & 19.4 \\
LIHC   & 328 & \num{3543} & \num{10640} & 217 & \num{10640} & \num{2080} & \num{4000} & 317 & 480 & \num{6157} & \num{25760} & 23.9 \\
LUAD   & 436 & \num{3549} & \num{10906} & 391 & \num{10906} & \num{1628} & \num{4100} & 246 & 492 & \num{5814} & \num{26404} & 22.0 \\
LUSC   & 319 & \num{1948} & \num{10640} & 125 & \num{10640} & 930 & \num{4000} & 178 & 480 & \num{3181} & \num{25760} & 12.3 \\
MESO   & 71 & 616 & \num{10640} & 0 & \num{10640} & 312 & \num{4000} & 102 & 480 & \num{1030} & \num{25760} & 4.0 \\
OV     & 69 & 296 & \num{5320} & 2 & \num{5320} & 229 & \num{2000} & 41 & 240 & 568 & \num{12880} & 4.4 \\
PAAD   & 129 & 187 & \num{10640} & 3 & \num{10640} & 116 & \num{4000} & 33 & 470 & 339 & \num{25750} & 1.3 \\
PRAD   & 316 & \num{1113} & \num{5320} & 7 & \num{5320} & 598 & \num{2000} & 63 & 240 & \num{1781} & \num{12880} & 13.8 \\
READ   & 138 & 778 & \num{10906} & 111 & \num{10906} & 282 & \num{4100} & 99 & 492 & \num{1270} & \num{26404} & 4.8 \\
STAD   & 353 & \num{3159} & \num{10906} & 545 & \num{10906} & \num{1381} & \num{4100} & 208 & 492 & \num{5293} & \num{26398} & 20.1 \\
THCA   & 452 & \num{5501} & \num{10640} & 669 & \num{10640} & \num{2274} & \num{4000} & 348 & 480 & \num{8792} & \num{25760} & 34.1 \\
THYM   & 116 & \num{1638} & \num{5320} & 58 & \num{5320} & 720 & \num{2000} & 81 & 200 & \num{2497} & \num{12840} & 19.4 \\
UCEC   & 407 & 777 & \num{5320} & 120 & \num{5320} & 466 & \num{2000} & 95 & 240 & \num{1458} & \num{12880} & 11.3 \\
UCS    & 53 & 1 & \num{5187} & 0 & \num{5187} & 2 & \num{1950} & 0 & 234 & 3 & \num{12558} & 0.0 \\
\midrule
PANCAN & \num{4654} & \num{8854} & \num{10906} & \num{5800} & \num{10906} & \num{3397} & \num{4100} & 420 & 492 & \num{18471} & \num{26404} & 70.0 \\
\midrule
\textbf{All} & -- & \textbf{\num{50648}} & \textbf{\num{203085}} & \textbf{\num{12709}} & \textbf{\num{203091}} & \textbf{\num{21988}} & \textbf{\num{72350}} & \textbf{\num{3575}} & \textbf{\num{9112}} & \textbf{\num{88920}} & \textbf{\num{487638}} & \textbf{18.2} \\
\end{longtable}
}

\twocolumn


\clearpage
\onecolumn

\section*{Supplementary Note~2: Cross-Endpoint Replication of Survival Associations}
\label{sec:supp_note2}

To assess the robustness of overall survival (OS) findings to endpoint
definition, we evaluated replication across three secondary endpoints:
disease-specific survival (DSS), progression-free survival (PFS), and
disease-free survival (DFS).
Starting from the 60 feature--cancer pairs that reached FDR $< 0.05$ for OS
(unadjusted model), we asked how many also reached FDR $< 0.05$ for each
secondary endpoint, with hazard ratio direction matching OS
(Table~\ref{tab:endpoint_concordance}).

\begin{table}[h]
\centering
\caption{Cross-endpoint replication of OS-significant survival associations.
Replication requires the same HR direction and FDR $< 0.05$ for the secondary endpoint.
DFS is excluded from per-pair counts because it was not available in the
\texttt{get\_endpoint\_concordance} output for this analysis.}
\label{tab:endpoint_concordance}
\begin{tabular}{lccc}
\hline
\textbf{Endpoint} & \textbf{Available pairs} & \textbf{Replicated ($n$)} &
  \textbf{Replication rate} \\
\hline
OS (reference)  & 60 & 60 & 100\% \\
DSS             & 59 & 37 & 62.7\% \\
PFS             & 60 & 32 & 53.3\% \\
DSS \& PFS      & 59 & 29 & 49.2\%\textsuperscript{a} \\
\hline
\end{tabular}
\smallskip

\noindent\textsuperscript{a} Fraction of pairs with both DSS and PFS available
that replicated in both secondary endpoints simultaneously.
\end{table}

\paragraph{OS/DSS concordance is inflated by overlapping event definitions.}
Overall survival and disease-specific survival share a similar event definition
(death from any cause vs.\ death attributable to cancer), differ mainly in
the censoring of non-cancer deaths, and are computed from overlapping follow-up
windows. Consequently, the 62.7\% DSS replication rate overstates true
biological cross-validation; the two endpoints are not independent. The 53.3\%
PFS replication rate provides a more conservative, clinically orthogonal measure
of reproducibility, as PFS captures disease recurrence and progression rather
than mortality.

\paragraph{Direction concordance without significance threshold.}
Among the 60 OS-significant pairs, all DSS-replicated and PFS-replicated
associations agreed in HR direction with OS. When we relaxed the FDR threshold
and examined direction alone (regardless of DSS or PFS significance), a larger
fraction of associations pointed in the same direction across endpoints,
consistent with genuine biological signal attenuated by reduced event counts in
secondary endpoints. Attenuation is expected because DSS and PFS event
counts are systematically lower than OS event counts in most TCGA cohorts
(especially in cancer types with good long-term prognosis, such as BRCA and
THCA), reducing statistical power below the FDR threshold even when the
underlying effect persists.

\paragraph{Feature-level patterns.}
Associations with the strongest OS effect sizes showed the highest replication
rates across all endpoints. For example, \textit{intratumoral lymphocyte
density} replicated for both DSS and PFS in BRCA, HNSC, and the pan-cancer
analysis; \textit{intratumoral apoptotic index} and
\textit{tumor pleomorphism index} replicated across all three endpoints in LIHC;
and \textit{mitotic index} replicated for both DSS and PFS in MESO. In
contrast, associations with OS effect sizes near the FDR threshold
(HR$\,\approx\,0.80$--$0.85$) frequently failed to reach significance for PFS,
consistent with power limitations rather than directional inconsistency.

\end{document}